\documentclass[a4paper,11pt]{article}
\pdfoutput=1 
\usepackage{jheppub}
\usepackage[T1]{fontenc} 
\usepackage{amsmath}
\usepackage{amssymb}
\usepackage{amsfonts}
\usepackage{mathtools}
\usepackage{microtype}
\usepackage{tikz}
\usepackage{bm}
\usepackage[utf8]{inputenc}
\usepackage{blkarray}
\usepackage{bigstrut}
\usepackage{nccmath}
\usepackage{enumitem}
\usepackage{braket}
\usepackage{dsfont}
\usepackage{lipsum}
\usepackage{array,multirow}
\usepackage{physics}
\usepackage{xcolor}
\usepackage{color}
\usepackage{float}
\usepackage[export]{adjustbox}
\usepackage{graphicx}
\usepackage{caption}
\usepackage{comment}
\usepackage{lscape}
\usepackage{bbm}
\usepackage{subcaption}
\usepackage{hyperref}

\allowdisplaybreaks

\usepackage{xargs}
\usepackage[colorinlistoftodos,prependcaption,textsize=tiny]{todonotes}

\usepackage{paracol}
\usepackage{rotating}
\usepackage{makecell}
\usepackage{longtable}
\usepackage{booktabs}
\usepackage{multirow}
\newcommand{\tabnum}[1]{\parbox{\widthof{\texttt{+3.405963008295366e-2*{\rm i}}}}{\texttt{#1}}}
\newcommand{\tworow}[1]{\multirow{2}{*}{\parbox{\widthof{#1}}{\centering #1}}}

\usepackage{nicematrix}

\newcommand{\Log}{\text{Log}}
\newcommand\numthis{\addtocounter{equation}{1}\tag{\theequation}}
\newcommand{\Line}{\textsc{LINE}}
\newcommand{\negq}{\mkern-16mu}

\newcolumntype{M}[1]{>{\centering\arraybackslash}m{#1}}
\newcolumntype{N}[1]{>{\centering\arraybackslash$}m{#1}<{$}}
\newcolumntype{C}[1]{>{\centering\arraybackslash}p{#1}}
\newcommand{\widesim}[2][1.5]{
  \mathrel{\overset{#2}{\scalebox{#1}[1]{$\sim$}}}
}

\newcommand{\off}{\text{off}}

\newcommand{\unipd}{Dipartimento di Fisica e Astronomia, Universit\`a degli Studi di Padova, and INFN, Sezione di Padova,\\ 
Via Marzolo 8, I-35131 Padova, Italy.}

\newcommand{\napoliall}{Dipartimento di Fisica, Universit\`a di Napoli Federico II, and INFN, Sezione di Napoli,\\
Via Cinthia, I-80126 Napoli, Italy.}

\setlist[itemize,1]{leftmargin=1.5em}

\title{LINE: Loop Integrals Numerical Evaluation}

\author[a]{Renato Maria Prisco,}
\author[b]{Jonathan Ronca,}
\author[a]{Francesco Tramontano}

\affiliation[a]{\napoliall}
\affiliation[b]{\unipd}

\emailAdd{renatomaria.prisco@unina.it}
\emailAdd{jonathan.ronca@pd.infn.it}
\emailAdd{francesco.tramontano@unina.it}

\preprint{-}

\abstract{
  We present methods for the numerical evaluation of the master integrals that appear in the calculation of scattering amplitudes at higher order in perturbative quantum field theory.
  We follow the general strategy of solving first-order ordinary differential equations through series expansion.
  We have collected these procedures in an open source computer program that we dub \Line{}.
  Boundary conditions can be provided by the user or computed internally using the method of expansion by regions.
  Illustrative examples are also given.
}

\begin{document} 

\maketitle

\section{Introduction}
\label{sec:introduction}

The success of the physics program of the Large Hadron
Collider at CERN paved the way for major new challenges to the
Standard Model.
The near-future collider experiments, like High Luminosity Large
Hadron Collider and the Future Circular electron-positron
Collider, aims at measuring physical observables at the
per-mille precision level over wide phase-space regions.
On the theoretical side, the core elements of every higher-order perturbative computation in quantum field theory are Feynman amplitudes.
Once expressed in terms of form factors, helicity structures, or direct interferences, they give rise to Lorentz-scalar integrals known as Feynman Integrals (FIs).

To match the precision of future experiments it is necessary to go beyond Leading Order (LO) or  Next-to-Leading Order (NLO) corrections.
Reliable comparisons with experimental data necessitate the inclusion of Next-to-Next-to-Leading Order (NNLO) and Next-to-Next-to-Next-to-Leading Order ($\mathrm{N}^3\mathrm{LO}$) contributions~\cite{Caola:2022ayt}.
Typically, thousands of two-loop and three-loop integrals contribute to NNLO and $\mathrm{N}^3$LO amplitudes. The computational complexity of these integrals increases rapidly with the number of loops, external legs, and physical energy scales. Noteworthy advancements include the computation of two-loop, five-point NNLO QCD amplitudes~\cite{Badger:2024sqv,Agarwal:2023suw} and some complete three-loop, up-to-three-point $\mathrm{N}^3$LO results~\cite{Bonetti:2017ovy,Fael:2022rgm,Forner:2024ojj,Duhr:2024bzt}. Regarding three-loop, four- and five-point amplitudes, significant progress has been made in computing many FIs~\cite{Henn:2023vbd,Gehrmann:2024tds,Long:2024bmi}. For higher-loop cases, results are available for relevant two-point amplitudes~\cite{Laporta:2017okg,Aoyama:2017uqe,Volkov:2019phy} and vacuum integrals~\cite{Luthe:2015ngq}.

FIs fulfill Integration-by-Parts Identities
(IBPs)~\cite{Tkachov:1981wb,Chetyrkin:1981qh,Laporta:2000dsw} that allow to express the integrals of a given amplitude as linear combination of a minimal set of independent Master
Integrals (MIs).
Over the past decades, numerous techniques have been developed to tackle the computation of Master Integrals (MIs). A wide range of analytical methods has been utilized, including direct integration via specific integral representations~\cite{Davydychev:1992mt,Usyukina:1992jd}, Difference Equations~\cite{Tarasov:1996br,Laporta:2000dsw}, and the Differential Equations (DEs) method~\cite{Barucchi:1973zm,Kotikov:1990kg,Kotikov:1991pm,Bern:1992em,Remiddi:1997ny,Gehrmann:1999as,Henn:2014qga,Argeri:2014qva}.
In addition, numerical techniques have been developed and can be broadly classified into two main categories: Monte Carlo-based approaches, such as sector decomposition~\cite{Binoth:2000ps,Heinrich:2008si,Borowka:2015mxa,Borowka:2017idc,Heinrich:2023til}, Mellin-Barnes techniques~\cite{Boos:1990rg,Smirnov:1999gc,Tausk:1999vh,Czakon:2005rk,Gluza:2007rt}, and tropical integration~\cite{Borinsky:2023jdv};
and numerical solution of DEs~\cite{Liu:2017jxz,Liu:2022chg,Moriello:2019yhu,Hidding:2020ytt,Armadillo:2022ugh}.
Notably, these methods have been successfully applied to several two- and three-loop calculations, including NLO EW corrections for $gg \to HH$~\cite{Bi:2023bnq}, NLO QCD corrections for $pp \to Hj$~\cite{Bonciani:2022jmb}, and other processes, see~\cite{Armadillo:2024nwk,Liu:2024ont,Chen:2022vzo} for examples.

The power of the DE method lies in the possibility to efficiently build high accuracy results.
In this paper we present \Line{} (which stands for Loop Integral Numerical Evaluation),
a novel tool to compute MIs by solving numerically DEs via series expansion.
\Line{} is mostly written in \texttt{C} and leverages the well-known \texttt{GMP} family of libraries for arbitrary precision arithmetic, aiming to achieve efficiency and accessibility in order to go beyond proof of concept and make large-scale cluster computations more feasible.
On general ground, the dependence on the space-time dimension can be treated in different ways.
One approach consists in expanding the DE matrices in the dimensional regularization parameter up to the desired order and solving for the expansion coefficients~\cite{Moriello:2019yhu,Hidding:2020ytt,Armadillo:2022ugh}.
However, we follow the alternative strategy of solving the DEs assigning several numerical values to the dimensional parameter to compute the Laurent series coefficients via interpolation~\cite{Liu:2017jxz,Liu:2022chg}.
This approach requires the knowledge at all orders in $\epsilon$ of boundary conditions for the MIs, that in \Line{} are computed through an implementation of the Auxiliary Mass Flow (AMFlow) method~\cite{Liu:2017jxz,Liu:2022chg}.
Such procedure relies on the introduction of an auxiliary squared mass and the use of Expansion By Regions (EBR)~\cite{Beneke:1997zp}.
Beyond that, in \Line{} we show a few examples where EBR can be used bypassing the introduction of an additional parameter, exploiting the singular structure of the DEs.

Furthermore, \Line{} implements techniques for the symbolic manipulation of DE matrices which are rational functions of the kinematic variables and the space-time dimension.
For instance, we carry out the transformation of the DE matrices to normalized Fuchsian form which is an essential step in the computation of the solution around regular singular point of the DEs.
In addition, algorithms to ensure the proper analytic continuation are implemented to solve around branch points of the MIs.




This paper is organized as follows.
In Section~\ref{sec:theory} we introduce the notation and the basics
of the method of DEs for MIs.
In Section~\ref{sec:method} we describe in detail the algorithms and techniques employed in \Line{} while in Section~\ref{sec:examples} we show applications for a selection of one- and two-loop examples.
Finally, in Appendix~\ref{sec:matr-norm} we discuss the transformation of the DEs to normalized Fuchsian form and in Appendix~\ref{sec:implementation} we give a few implementation details.

\section{Feynman Integrals}
\label{sec:theory}
The central quantities in this work are the (dimensionally regularized) Feynman integrals
\begin{equation}\label{eq:fidef}
  F^{\boldsymbol{\nu}}(\epsilon, {\bf s}) = \int_{k_1,\dots,k_l} \prod_{\alpha=1}^n\frac{1}{D_\alpha^{\nu_\alpha}}, \qquad D_\alpha = q_\alpha^2 - m_\alpha^2 + i\varepsilon \,,
\end{equation}
where $\epsilon$ is the dimensional regularization parameter, ${\bf s}$ is a vector of invariant mass-scales (which includes the internal masses $m_\alpha$), ${\boldsymbol{\nu}}\equiv\{\nu_\alpha\}$ is the integer-valued vector of powers $\nu_\alpha$ of the inverse propagators $D_\alpha$, $q_\alpha$ is a linear combination of loop and external momenta with coefficients $\pm 1$, $\varepsilon$ is a small positive imaginary part that enforces the Feynman prescription,
$l$ is the number of loops and $n$ is the number of propagators. The integral notation reads
\begin{equation}
  \int_{k_1,\dots,k_l} = \int\prod_{i=1}^l \frac{d^d k_i}{i\pi^{\frac{d}{2}}} \,,
\end{equation}
which is related to the physical FIs normalization as follows:
\begin{equation}
   \mu^{4-d}\int\prod_{i=1}^l \frac{d^d k_i}{(2\pi)^{d}} = \frac{i}{(4\pi^2)}(4\pi \mu^2)^{2-\frac{d}{2}}\int_{k_1,\dots,k_l} \,.
\end{equation}
Here, $d=4-2\epsilon$ is the space-time dimension.

The set of denominators $D_\alpha$ defines an {\it integral family}. FIs with different $\boldsymbol{\nu}$ but same set of $D_\alpha$ belong to the same integral family.
The subset $\mathcal{T}_{F^{\boldsymbol{\nu}}} \equiv \{D_\alpha | \nu_\alpha>0 \}$ of denominators with positive powers is called {\it topology} of $F^{\boldsymbol{\nu}}(\epsilon, {\bf s})$, which is related to the topology of the associated Feynman graph; an integral $F^{\boldsymbol{\nu}'}(\epsilon, {\bf s})$ with topology $\mathcal{T}_{F^{\boldsymbol{\nu}'}}\subset \mathcal{T}_{F^{\boldsymbol{\nu}}}$ is said to belong to a {\it sub-topology} of $F^{\boldsymbol{\nu}}(\epsilon, {\bf s})$.

\subsection{Integration-by-parts identities}
Feynman integrals of the form in Eq.\,\eqref{eq:fidef} satisfy the following set of identities:
\begin{equation}\label{eq:ibps}
  \int_{k_1,\dots,k_l} \frac{\partial}{\partial k_i^\mu}\left(v^\mu_j \prod_{\alpha=1}^n\frac{1}{D_\alpha^{\nu_\alpha}}\right) = 0 \,.
\end{equation}
where $v_j$ can be for example a loop momentum, an external momentum or a linear combination of them.
These relations are the well-known {\it integration-by-parts identities} \cite{Tkachov:1981wb,Chetyrkin:1981qh,Laporta:2000dsw}.

IBPs introduce linear relations between integrals $F^{\boldsymbol{\nu}}(\epsilon, {\bf s})$ with different powers $\boldsymbol{\nu}$ but belonging to the same integral family.
For a given family, it can be shown that there exists a finite set of integrals ${\bf I}(\epsilon, {\bf s})$, called {\it master integrals}, such that every other $F^{\boldsymbol{\nu}}(\epsilon, {\bf s})$ can be written as a linear combination of MIs:
\begin{equation}
 F^{\boldsymbol{\nu}}(\epsilon, {\bf s}) = {\bf C}({\boldsymbol{\nu}}; \epsilon, {\bf s}) \cdot  {\bf I}(\epsilon, {\bf s}) \,,
\end{equation}
where ${\bf C}({\boldsymbol{\nu}}; \epsilon, {\bf s})$ is a vector of rational functions of the scales ${\bf s}$ and the dimensional regulator~$\epsilon$. 

The standard techniques to find the number of independent MIs and choose a basis rely on the generation of a linear system of equation by varying $k_i$, $v_j$ and the indices $\nu_\alpha$ in the IBPs relations of Eq.\,\eqref{eq:ibps}.
After counting the number of independent equations, the choice of MIs can be done algorithmically~\cite{Laporta:2000dsw}.
The linear system is then solved by Gauss elimination, which yields the specific linear combination of MIs for any integral $F^{\boldsymbol{\nu}}(\epsilon, {\bf s})$.
Tools that implements this techniques can be found in~\cite{Studerus:2009ye,Lee:2012cn,Maierhofer:2017gsa,Guan:2024byi}.

\subsection{Differential equations for master integrals}
\label{sec:DEs-for-MIs}
The IBPs can be exploited to build a system of differential equations for the MIs~\cite{Barucchi:1973zm,Kotikov:1990kg,Kotikov:1991pm,Bern:1992em,Remiddi:1997ny,Gehrmann:1999as,Henn:2014qga,Argeri:2014qva}.
These are functions of the dimensional shift parameter and the kinematic invariants ${\bf s}$ and their derivatives with respect to the latter can be written as linear combinations of FIs $F^{\boldsymbol{\nu}_{jm}}(\epsilon, {\bf s})$ with different indices ${\boldsymbol \nu}_m$:
\begin{equation}
  \frac{\partial}{\partial s_i}I_j(\epsilon, {\bf s}) = \sum_{m} C'_{ijm}(\epsilon, {\bf s}) F^{\boldsymbol{\nu}_{jm}}(\epsilon, {\bf s}) \,.
\end{equation}
In addition, IBPs give linear relations between $F^{\boldsymbol{\nu}_{jm}}(\epsilon, {\bf s})$ and the MIs ${\bf I}(\epsilon, {\bf s})$, so that
\begin{equation}
  F^{\boldsymbol{\nu}_{jm}}(\epsilon, {\bf s}) = {\bf C}(\boldsymbol{\nu}_{jm}; \epsilon, {\bf s}) \cdot  {\bf I}(\epsilon, {\bf s})\quad \implies \quad \frac{\partial}{\partial s_i}I_j(\epsilon, {\bf s}) = \sum_{m} C'_{ijm}(\epsilon, {\bf s}) {\bf C}(\boldsymbol{\nu}_{jm}; \epsilon, {\bf s}) \cdot  {\bf I}(\epsilon, {\bf s}) \,,
\end{equation}
which can be written in the more suggestive form
\begin{equation}\label{eq:gendemulti}
  \frac{\partial}{\partial s_i}{\bf I}(\epsilon, {\bf s}) = \mathbb{A}_i(\epsilon, {\bf s}){\bf I}(\epsilon, {\bf s}) \,,
\end{equation}
or equivalently, as a total differential
\begin{equation}
  d{\bf I}(\epsilon, {\bf s}) = d\mathbb{A}(\epsilon, {\bf s}){\bf I}(\epsilon, {\bf s}) \,, \quad d\mathbb{A}(\epsilon, {\bf s}) = \sum_i \mathbb{A}_i(\epsilon, {\bf s})ds_i \,.
\end{equation}
The entries of the DE matrices $\mathbb{A}_i(\epsilon, {\bf s})$ are then rational functions of $\epsilon$ and ${\bf s}$.



Solving Eq.\,\eqref{eq:gendemulti} requires the knowledge of $\mathbf{I}(\epsilon, {\bf s})$ at a specific phase-space point ${\bf s}_0$.
Finding boundary conditions is, in general, a non-trivial task.
A common approach involves imposing the regularity of $\mathbf{I}(\epsilon, \mathbf{s})$ at specific kinematical points, such as non-physical thresholds, where the integrals become significantly easier to evaluate. Typically, these points correspond to simple integral families, like vacuum or bubble integrals, and possibly their products.
As a reference example, consider the boundary of a one-loop bubble integral
\begin{equation} \includegraphics[width=.2\textwidth,valign=c]{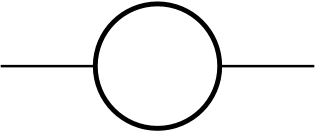} = \int_k \frac{1}{[k^2 - m^2][(k+p)^2 - m^2]} \,,
\end{equation}
which can be determined by taking the limit $p \to 0$
\begin{equation}
  \lim_{p\to 0} \includegraphics[width=.2\textwidth,valign=c]{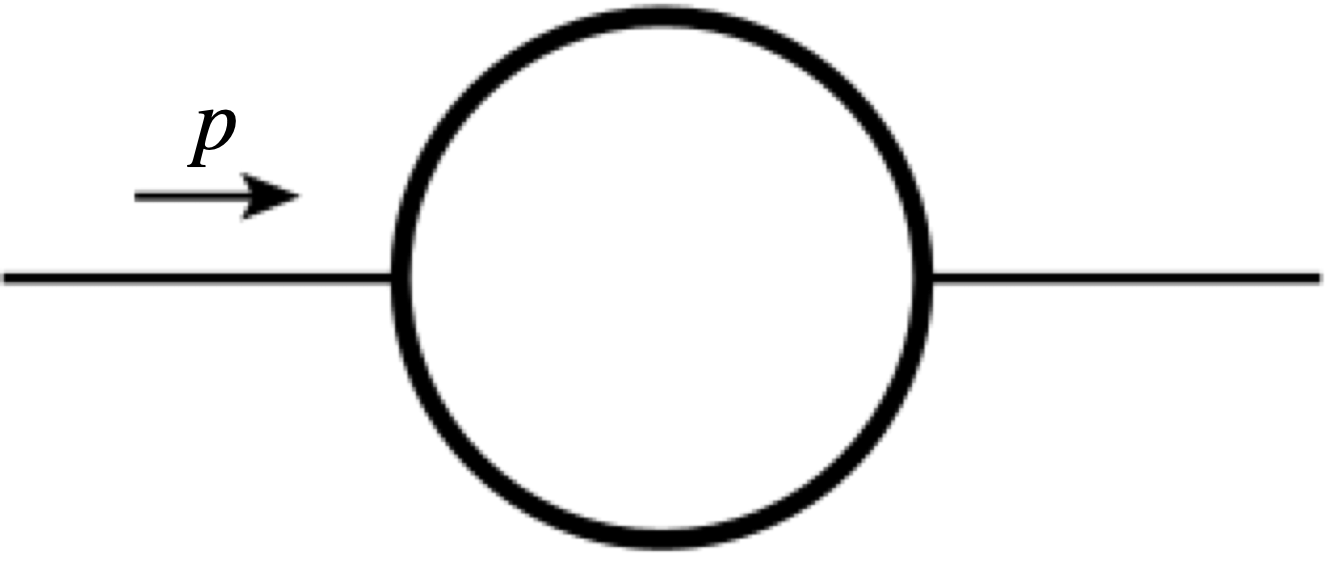} \,\,= \, \int_k \frac{1}{[k^2 - m^2]^2} =\includegraphics[width=.085\textwidth,valign=c]{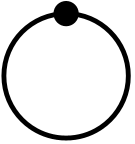} =  \frac{\Gamma(1+\epsilon)}{\epsilon}(m^2)^{-\epsilon}  \,.
\end{equation}
This result corresponds to the well-known tadpole integral, for which an exact expression in terms of its parameters is readily available in the literature.

The DE method for the evaluation of MIs has a long history of successes, and has been employed to achieve analytical evaluation of FIs in terms of special functions, like Generalized polylogarithms, Harmonic polylogarithms, Pentagon functions and other special functions~\cite{Remiddi:1999ew,Goncharov:1995tdt,Gehrmann:2015bfy,Gorges:2023zgv}. State-of-the-art calculations exploiting this method cover classes of FIs with up to three-loop and five-scales~\cite{Liu:2024ont}. Unfortunately, this approach is not fully algorithmic. A more systematic method can be defined for matrices in the so-called {\it canonical form} \cite{Henn:2013pwa} and introducing the notion of {\it letters}, which better encode the analytic properties of FIs.
For a more detailed discussion about the latter technique, we refer to \cite{Henn:2013pwa,Henn:2014qga,Zoia:2021zmb} and references therein.

Numerical solutions of DEs for MIs have recently gained interest. The idea is the following: given the matrices $\mathbb{A}_i(\epsilon, {\bf s})$ and the boundary conditions $\mathbf{I}(\epsilon, {\bf s}_0)$, one can solve the DEs by means of series expansion in the kinematical variables, whose coefficients are fixed by the inputs; such a solution is then evaluated to another phase-space point and its value can be used as a new boundary for further propagations to reach any target point.
In the last few years, several tools that implement this method have been proposed~\cite{Moriello:2019yhu,Liu:2022chg,Hidding:2020ytt,Armadillo:2022ugh}. 


In this work, we present in detail a strategy for solving DEs numerically, which is implemented in \Line{}.
In our framework, to propagate the MIs from a given boundary point ${\bf s}_0$ to a target ${\bf s}_1$, we build the associated univariate equation on the straight line connecting ${\bf s}_0$ and ${\bf s}_1$
\begin{equation}
  \label{eq:line}
  {\bf s}(\eta) = {\bf s}_0 + ({\bf s}_1 - {\bf s}_0)\eta \,,\quad \eta\in[0,1] \,,
\end{equation}
through the introduction of a line parameter $\eta$. The differential form $d\mathbb{A}(\epsilon, {\bf s})$ transforms as follows:
\begin{equation}
   d\mathbb{A}(\epsilon, {\bf s}) = \left[\sum_i \mathbb{A}_i(\epsilon, {\bf s})\left(\frac{\partial s_i}{\partial \eta}\right)\right]d\eta = \mathbb{A}(\epsilon, \eta)d\eta \,,
\end{equation}
so that the DE w.r.t. the line parameter $\eta$ becomes
\begin{equation}\label{eq:univariatede}
  \frac{d}{d\eta}{\bf I}(\epsilon, \eta) = \mathbb{A}(\epsilon, \eta){\bf I}(\epsilon, \eta) \,, \quad {\bf I}(\epsilon, 0) = {\bf I}_{bc}(\epsilon) \,.
\end{equation}

To evaluate the coefficients of the Laurent series
\begin{equation}
  {\bf I}(\eta, \epsilon) = \epsilon^{-2l}\sum_{j=0}^{\texttt{ord}} {\bf I}_j(\eta)\epsilon^j + O(\epsilon^{-2l+{\texttt{ord} }+1})
\end{equation}
up to the desired order $-2 l + \texttt{ord}$, we follow the same strategy as \cite{Liu:2022chg} and interpolate the solutions of the DEs for $N_{\epsilon}$ numerical values $\{\epsilon_0, \epsilon_1, \dots\}$ of the dimensional shift parameter $\epsilon$, where
\begin{align*}
  N_{\epsilon} &= \text{ceil} \left( 5 \frac{\texttt{ord}}{2} + 2 l \right) \,, \\
  \epsilon_k &= (101 + k) \times 10^{\{-\texttt{prec}/(\texttt{ord}+1) - l/2 - 2\}} \numthis \,,
\end{align*}
$\texttt{prec}$ being the required number of digits for the final result.
The above setup is the one chosen in~\cite{Liu:2022chg} and
has proven to be sufficient for all the applications we have studied with \Line{} so far.

Notice that, for fixed value of $\epsilon$, the propagation method is purely numerical. The relevant quantities are the (numerical) coefficients of the series expansions, as well as the coefficients of the polynomials appearing in the DE matrix $\mathbb{A}(\epsilon_k, \eta)$. In Appendix~\ref{sec:rat-func} we discuss the way rational functions are defined in our implementation.

The simplification of using numerical values for $\epsilon$ comes at the cost that boundary conditions must be known at all orders in $\epsilon$.
To this extend, we built an in-house implementation of the {\rm auxiliary-mass flow} method\footnote{In particular, we implemented the basic strategy proposed in~\cite{Liu:2017jxz}, where the auxiliary squared mass parameter is added to all denominators.}
to compute boundaries at arbitrary precision for any given phase-space point.
Details on our implementation are presented in Section~\ref{sec:amf}.

For each $\epsilon_k$, we solve the corresponding DEs in $\eta$ via series expansion.
Since the radius of convergence is finite, we split the line in Eq.\,\eqref{eq:line} in several segments following the strategy outlined in Section~\ref{sec:path-finder}.
We then propagate the boundary conditions through the subsequent construction of the series solution around the endpoints of each segment.
The expansions points are either regular or regular singular points of the DEs and in Section~\ref{sec:method} we describe the mathematical procedure to build the solution around both kind of points.

\section{General methods of series expansion solution to DE}
\label{sec:method}
\label{sec:seriessolde}
Master integrals are analytic functions and so admit series
representation. In what follows we present strategies to find the
series coefficients up to a an assigned integer power
$\omega$ of the solution to the DEs
\begin{equation}\label{eq:eqdiff-full}
    \frac{d}{d\eta}{\mathbf I}(\epsilon, \eta) = {\mathbb A}(\epsilon, \eta){\mathbf
      I}(\epsilon, \eta) \,,\quad {\mathbf I}(\epsilon, \eta_{bc}) = {\mathbf I}_{bc}(\epsilon) \,.
\end{equation}
From now on, we implicitly assume a fixed value of $\epsilon$, so that Eq.\,\eqref{eq:eqdiff-full} becomes univariate. Consequently, we omit $\epsilon$ from the arguments of the functions.

The algorithm to find a series solution for a system of Differential Equations (DEs) is quite simple in principle.
The first step consists in writing an appropriate ansatz for the solution as a series in the line parameter, with unknown coefficients to be substituted into the DEs.
This leads to recurrence relations among the unknown coefficients.
Next, the boundary values of the solution are used to fully determine the series expansion.
Along the path connecting the boundary and target points, poles of the DEs may be encountered.
To handle such singularities, we can construct a local solution around the singular point by introducing a suitable ansatz for the series.
This ansatz is designed to capture the specific singular behavior of the solution near the singularity.

In this Section we describe step by step how to split into segments the line connecting the initial and final points, build the solution around regular and singular points, and perform analytic continuation when crossing a singularity.

\subsection{Path finder}\label{sec:path-finder}
The presence of poles in the DE matrix causes the series expansion of the solution to have a finite radius of convergence equal to the distance from the closest pole.
For this reason, the propagation of the MIs from the initial to the target point is split into multiple steps by placing a set of points along the phase-space line so that the radius of convergence at any point is bigger than the distance from the next point.

In \Line{} this is done by first identifying the singular points $\sigma_i$ along the phase-space line; then, for every point $\sigma_i$ the radius of convergence $\mathcal{R}(\sigma_i)$ is computed and two \textit{matching points} $\mu_i^L, \mu_i^R$ are placed on the left and on the right of $\sigma_i$, respectively, according to
\begin{eqnarray}
  \mu_i^L & = & \sigma_i - \frac{\mathcal{R}(\sigma_i)}{2} \nonumber \,, \\
  \mu_i^R & = & \sigma_i + \frac{\mathcal{R}(\sigma_i)}{2} \,.
\end{eqnarray}
The choice of solving within half the radius of convergence makes analytic continuation easier, as discussed in Section~\ref{sec:analyt-cont}.

If the MIs $\mathbf{I}(\eta)$  have been propagated up to the matching point $\mu_i^L$, the DEs can be solved around the singular point $\sigma_i$ using the value $\mathbf{I}(\mu_i^L)$ as boundary condition and the solution can be evaluated in $\mu_i^R$.
To propagate from $\mu_i^R$ to the matching point $\mu_{i+1}^L$ on the left of the next singular point $\sigma_{i+1}$, a set of regular points $\rho_i^j$ is found by moving along the phase-space line with steps given by half the radius of convergence at the current point
\begin{eqnarray}
\rho_i^0 & = & \mu_i^R + \frac{\mathcal{R}(\mu_i^R)}{2} \,, \nonumber \\
\rho_i^{k+1} & = & \rho_i^k + \frac{\mathcal{R}(\rho_i^k)}{2} \,, \quad k \ge 1 \,,
\end{eqnarray}
until, for some $K$, the condition $\mu_{i+1}^L \le \rho_i^K + \mathcal{R}(\rho_i^K)/2$ is satisfied.

The regular points between $\eta = 0$ and the first singular point, as well as those from the last singular point to $\eta = 1$, are found in an analogous way, considering $\eta = 0$ and $\eta = 1$ as matching points.

The set of points
\begin{equation}
\mathcal{P} = \{ 0, \rho_0^1, \rho_0^2, \dots , \mu_1^L, \sigma_1, \mu_1^R, \rho_1^1, \rho_1^2, \dots, \mu_2^L, \sigma_2, \mu_2^R, \rho_2^1, \rho_2^2, \dots, 1\}
\end{equation}
is referred to as the \textit{phase-space path} between the starting and the target point.

We note that alternative strategies have been proposed in the literature, see for example~\cite{Frellesvig:2019byn}.

\subsection{Solving DE around regular points}
Consider a set of master integrals related by the first order
(univariate) differential equation
\begin{equation}\label{eqdiff}
    \frac{d}{d\eta}{\mathbf I}(\eta) = {\mathbb A}(\eta){\mathbf
      I}(\eta) \,, \quad {\mathbf I}(\eta_{bc}) = {\mathbf I}_{bc} \,.
\end{equation}
For any regular point $\eta_{bc}$, ${\mathbb A}(\eta_{bc})$ is not singular,
and in a neighborhood of $\eta_{bc}$ Eq.\,\eqref{eqdiff} admits a solution
whose Taylor series can be written as
\begin{equation}
    {\mathbf I}(\eta) = \sum_{j=0}^\infty {\mathbf I}_j(\eta -
    \eta_{bc})^j \,.
\end{equation}

In order to simplify the notation, we shift the matrix ${\mathbb A}(\eta)$ by $\eta \to \eta - \eta_{bc}$ so that
the sought solution takes the form
\begin{equation}\label{regsol}
    {\mathbf I}(\eta) = \sum_{j=0}^\infty {\mathbf I}_j\eta ^j \,,
\end{equation}
while the matching point becomes $\eta = 0$.
Since ${\mathbf I}(0) = {\mathbf I}_0$, the boundary conditions is satisfied by setting ${\mathbf I}_0 = {\mathbf I}_{bc}$.

Ordering the MIs so that simplest sub-topologies appear first, the matrix ${\mathbb A}(\eta)$ results to be in lower-triangular form. Therefore,
one can identify a block structure and solve the non-homogeneous
differential equations for each block sequencially. The differential equation for the
$r$-th block is
\begin{equation}\label{eq:nheq}
    \frac{d}{d\eta}{\mathbf I}^{(r)}(\eta) = {\mathbb
      A}^{(r)}(\eta){\mathbf I}^{(r)}(\eta) + {\mathbf J}^{(r)}(\eta) \,,
\end{equation}
where ${\mathbb A}^{(r)}$ is the $r$-th diagonal block of ${\mathbb A}$, ${\mathbf J}^{(r)}(\eta)$ is build contracting the $r$-th block row of ${\mathbb A}(\eta)$ with the solution of the previous $r-1$ blocks and expanding around $\eta=0$.
We define
\begin{equation}
  \begin{gathered}
    g^{(r)}(\eta) = {\rm
      LCD}\left(\mathbb{A}^{(r)}(\eta)\right) \,, \quad
    \mathbb{B}^{(r)}(\eta) =
    g^{(r)}(\eta)\mathbb{A}^{(r)}(\eta)\,,\\ {\mathbf Y}^{(r)}(\eta)
    = g^{(r)}(\eta){\mathbf J}^{(r)}(\eta) \,.
  \end{gathered}
\end{equation}
From now on, we refer all the equations to a single block $r$ and for an easier notation we omit such index.
One has
\begin{equation}\label{polydef}
  \begin{gathered}
    g(\eta) =\sum_{n=0}^\infty g_n\eta^n\,,\quad \mathbb{B}(\eta) =
    \sum_{m=0}^\infty \mathbb{B}_m\eta^m\,, \\ \mathbf{Y}(\eta) =
    \sum_{m=0}^\infty \mathbf{Y}_m\eta^m\,,
  \end{gathered}
\end{equation}
where it is understood that the coefficients  $g_n$ and $\mathbb{B}_m$ are zero for $n>N, m>M$ (with $N$ and $M$ the polynomial degrees of $g^{(r)}(\eta)$ and $\mathbb{B}^{(r)}(\eta)$ respectively), and $\mathbf{Y}(\eta)$ is a Taylor expansion. Note that, being $\mathbb{A}(\eta)$ (and thus also $\mathbb{A}^{(r)}(\eta)$) non-singular for $\eta=0$, $g_0$ is non-vanishing.

By multiplying both sides of Eq.\,(\ref{eq:nheq}) times $g(\eta)$,
plugging the ansatz of Eq.\,(\ref{regsol}) and using the definitions in Eq.\,(\ref{polydef}), we obtain
\begin{equation}
    \sum_{n,k} k g_n {\mathbf I}_k\eta^{k+n-1} = \sum_{m, l}
    \mathbb{B}_m{\mathbf I}_l\eta^{l+m} + \sum_{r} \mathbf{Y}_r\eta^r\,.
\end{equation}
Moving the first term of the r.h.s. to the left and renaming the
indices conveniently, we have
\begin{equation}
    \sum_{n,k}\left[n g_{k+1-n} - \mathbb{B}_{k-n}\right]{\mathbf
      I}_{n}\eta^{k}= \sum_{k} \mathbf{Y}_k\eta^k\,.
\end{equation}
For any order $k$, one has
\begin{equation}
    \sum_{n=a(k)}^{k+1}\left[n g_{k+1-n} - \mathbb{B}_{k-n}\right]{\mathbf I}_{n}= \mathbf{Y}_k\,.
  \end{equation}
Since, by construction, $\mathbb{B}_{a<0} = 0$, we can easily isolate the last term of the sum
\begin{eqnarray}\label{regrec}
  (k+1) g_0 \mathbf{I}_{k+1} &&= \mathbf{Y}_k - \sum_{n=a(k)}^k \left[n g_{k+1-n} - \mathbb{B}_{k-n}\right]{\mathbf I}_{n} \nonumber \,,\\
  \mathbf{I}_{k+1}&&= \frac{1}{(k+1) g_0}\left[\mathbf{Y}_k - \sum_{n=a_1(k)}^k n g_{k+1-n} {\mathbf I}_{n}  +\sum_{n=a_2(k)}^k \mathbb{B}_{k-n} {\mathbf I}_{n}\right] \,,
\end{eqnarray}
where $a_1(k)= \max\{0,k+1-N\}$ and $a_2(k)= \max\{0,k-M\}$. 
Eq.\,(\ref{regrec}) is a recurrence relation that allows to compute any Taylor coefficient of $\mathbf{I}(\eta)$ starting from its boundary condition. We stress that the number of terms contributing to the two sums on the r.h.s. of Eq.\,(\ref{regrec}) is $N$ and $M-1$ respectively, $g(\eta)$ and $\mathbb{B}(\eta)$ being polynomials of finite degree.

\subsection{Solving DE around regular singular points}
Let us now consider Eq.\,(\ref{eqdiff}) around a pole of ${\mathbb A}(\eta)$. It is well known that singularities of differential equations for master integrals must be regular singular, therefore the solution admits a series expansion in their neighborhood. 
Without loss of generality, we shift the DE in such a way that $\eta=0$ is the regular singular point around which we expand the solution $\mathbf{I}(\eta)$.

We define the Poincaré rank of ${\mathbb A}(\eta)$ in $\eta=0$ as the lowest integer $p$ such that
\begin{equation}
  \lim_{\eta\to 0} \eta^{p+1}{\mathbb A}(\eta) = {\rm finite}\,.
\end{equation}
Being $\eta=0$ a regular singular point, it is possible to decrease the Poincaré rank $p$ of ${\mathbb A}(\eta)$ down to zero by applying a suitable transformation $\mathbb{T}(\eta)$.
The transformed DEs system reads:
\begin{equation}
  \label{eq:DE-fuch}
  \frac{d}{d\eta}\widetilde{\mathbf I}(\eta) = \widetilde{\mathbb A}(\eta)\widetilde{\mathbf I}(\eta)\,, \quad \widetilde{\mathbf I}(\eta_{bc}) = \widetilde{\mathbf I}_{bc}\,.
\end{equation}
where
\begin{eqnarray}
  \label{eq:DE-transf}
  \widetilde{\mathbf{I}}(\eta) &&= {\mathbb T}^{-1}(\eta) \mathbf{I}(\eta)\,, \quad   \widetilde{\mathbf{I}}_{bc} = {\mathbb T}^{-1}(\eta_0) \mathbf{I}_{bc}\,, \nonumber \\
  \widetilde{\mathbb{A}}(\eta) &&= {\mathbb T}^{-1}(\eta) \mathbb{A}(\eta) {\mathbb T}(\eta) - {\mathbb T}^{-1}(\eta) \frac{d}{d\eta} {\mathbb T}(\eta)\,.
\end{eqnarray}
A method for constructing the transformation $\mathbb{T}(\eta)$ is discussed in \cite{Lee:2014ioa}, and its implementation in \Line{} is shown in Appendix~\ref{sec:matr-norm}.

The transformed matrix $\widetilde{\mathbb{A}}(\eta)$ is in the so-called {\it Fuchsian form}, i.e. it has Poincaré rank $p=0$ in $\eta=0$, and it has the Laurent expansion
\begin{equation}
  \widetilde{\mathbb{A}}(\eta) = \frac{1}{\eta} \sum_{k=0}^\infty\widetilde{\mathbb{A}}_k \eta^k\,.
\end{equation}
It is known that, in such a basis, a solution for Eq.\,(\ref{eq:DE-fuch}) around $\eta=0$ admits the following expansion:
\begin{equation}\label{eq:sol}
  \begin{aligned}
    {\mathbf I}(\eta)&=\sum_{\lambda\in S}\eta^\lambda \sum_{k=0}^{K_\lambda}\frac{\log^k(\eta)}{k!} \sum_{n=0}^{\infty}{\mathbf I}_{\lambda,k,n}\eta^n\,,
  \end{aligned}
\end{equation}
where $S$ is the set of eigenvalues of $\widetilde{\mathbb{A}}_0$, and $K_\lambda$ are the (finite) highest logarithmic powers associated with each eigenvalue $\lambda \in S$.

As for the regular case, we focus on the solution of a single block $(r)$, so that Eq.\,(\ref{eq:DE-fuch}) reads
\begin{equation}\label{eq:odezeroblock}
  \frac{d}{d\eta}{\mathbf I}^{(r)}(\eta) = \widetilde{\mathbb{A}}^{(r)}(\eta){\mathbf I}^{(r)}(\eta) + {\mathbf J}^{(r)}(\eta)\,.
\end{equation}
Furthermore, when building the transformation in Eq.\,\eqref{eq:DE-transf}, in our construction we also require the leading order of every block $\widetilde{\mathbb{A}}^{(r)}$ to be in \textit{Jordan normal form} and {\it free of resonances}, i.e. no pair of its eigenvalues has an integer difference.
To solve Eq.\eqref{eq:odezeroblock}, it is useful to define the following quantities:
\begin{equation}
  \begin{gathered}
    g^{(r)}(\eta) = {\rm LCD}\left(\eta\widetilde{\mathbb{A}}^{(r)}(\eta)\right) \,,
    \quad
    {\mathbb{B}}^{(r)}(\eta) = \eta g^{(r)}(\eta)\widetilde{\mathbb{A}}^{(r)}(\eta) \,,\\ 
    {\mathbf Y}^{(r)}(\eta)  = \eta g^{(r)}(\eta){\mathbf J}^{(r)}(\eta) \,,
  \end{gathered}
\end{equation}
and their series representation (omitting the block index $r$):
\begin{equation}\label{eq:polydef-sing}
  \begin{gathered}
    g(\eta) =\sum_{n=0}^\infty g_n\eta^n \,,\quad 
    {\mathbb{B}}(\eta) = \sum_{m=0}^\infty {\mathbb{B}}_m\eta^m \,, \\ 
    \mathbf{Y}(\eta) = \sum_{\lambda\in S}\eta^\lambda \sum_{k=0}^{K_\lambda}\frac{\log^k(\eta)}{k!} \sum_{m=0}^{\infty}{\mathbf Y}_{\lambda,k,m}\eta^m \,.
  \end{gathered}
\end{equation}
Once again, $g(\eta)$ and $\mathbb{B}(\eta)$ are polynomials of degree $N$ and $M$ respectively.
Also, we allow the coefficients ${\mathbf I}_{\lambda,k,n}$ and ${\mathbf Y}_{\lambda,k,n}$ to be zero for some $\lambda$.

By multiplying Eq.\,(\ref{eq:odezeroblock}) times $\eta g(\eta)$ we get
\begin{equation}
  \label{eq:de-block}
  \eta g(\eta) \frac{d}{d\eta}{\mathbf I}^{(r)}(\eta) = \widetilde{\mathbb{B}}^{(r)}(\eta){\mathbf I}^{(r)}(\eta) + {\mathbf Y}^{(r)}(\eta) \,.
\end{equation}
and using Eqs.\,(\ref{eq:polydef-sing}), we can express each term of the equation in its series representation:
\begin{eqnarray}\label{Ider}
  \eta g(\eta)\frac{d}{d\eta}{\mathbf I}(\eta) & = & \sum_{i,\lambda,k,n}  g_i\left[ {\mathbf I}_{\lambda,k+1,n-i}+(\lambda+n-i) {\mathbf I}_{\lambda,k,n-i}\right]\frac{\log^k(\eta)}{k!}\eta^{\lambda+n} \,,\\
  {\mathbb B}(\eta){\mathbf I}(\eta) & = & \sum_{j,\lambda,k,n}  {\mathbb B}_j{\mathbf I}_{\lambda,k,n-j}\frac{\log^k(\eta)}{k!}\eta^{\lambda+n} \,.
\end{eqnarray}
Then, for fixed powers $\lambda,n,k$ one has
\begin{equation}
  \label{eq:recsing-start}
  \sum_{i=0}^{\min\{N, n\}} \negq g_i {\mathbf I}_{\lambda,k+1,n-i} + \sum_{i=0}^{\min\{N,n\}} \negq (\lambda+n-i)g_i {\mathbf I}_{\lambda,k,n-i} 
  = \sum_{i=0}^{\min\{M,n\}} \negq {\mathbb B}_i {\mathbf I}_{\lambda,k,n-i} + {\mathbf Y}_{\lambda,k,n} \,.
\end{equation}
From Eq.\,\eqref{eq:recsing-start}, we can extract two recurrence relations:
\begin{align}
  {\mathbf I}_{\lambda,k+1,n} &= \nonumber \\
  \label{eq:rec-log}
  &\frac{1}{g_0} \left[ \sum_{i=0}^{\min\{M,n\}} \negq {\mathbb B}_i {\mathbf I}_{\lambda,k,n-i} - \sum_{i=0}^{\min\{N,n\}} \negq (\lambda+n-i)g_i {\mathbf I}_{\lambda,k,n-i} - \sum_{i=1}^{\min\{N,n\}} \negq  g_i {\mathbf I}_{\lambda,k+1,n-i} + {\mathbf Y}_{\lambda,k,n} \right] , \\
  {\mathbf I}_{\lambda,k,n} &= \left[ {\mathbb B}_0 - (\lambda+n)g_0 \mathbbm{1} \right]^{-1} 
\nonumber \\
  \label{eq:rec-eta}
  & \left[\sum_{i=0}^{\min\{N,n\}} \negq g_i {\mathbf I}_{\lambda,k+1,n-i} - \sum_{i=1}^{\min\{M,n\}} \negq {\mathbb B}_i {\mathbf I}_{\lambda,k,n-i} - \sum_{i=1}^{\min\{N,n\}} \negq (\lambda+n-i)g_i {\mathbf I}_{\lambda,k,n-i} - {\mathbf Y}_{\lambda,k,n} \right] \,.
\end{align}
In the following we illustrate how to exploit these formulae to build the sought series.

Note that $g_0\neq 0$ by construction.
Furthermore
\begin{equation}
  \label{eq:}
  {\mathbb B}_0 - (\lambda+n)g_0 \mathbbm{1} = g_0 \left[ {\mathbb A}_0 - (\lambda + n) \mathbbm{1} \right]
\end{equation}
is invertible if and only if $\lambda + n$ is not an eigenvalue of the matrix $\mathbb{A}_0$.
This is guaranteed by $n>0$ and the absence of resonances.

As usual, the solving procedure consists in building the general solution of the associated homogeneous equation as a linear combination of $L$ independent solutions with unconstrained coefficients, $L$ being the dimension of the block.
Then we add a particular solution of the non-homogeneous problem and fix the $L$ coefficients requiring that the boundary condition is satisfied.

\subsubsection{Solutions of the homogeneous equation}
\label{sec:homogeneous-solution}
We look for $L$ independent solutions of Eq.\,\eqref{eq:de-block} with ${\mathbf Y}^{(r)} = 0$, arranged as columns of a matrix ${\mathbb H}$ in the form
\begin{equation}
  {\mathbb H}(i,j)=\sum_{\lambda\in S}\eta^\lambda
  \sum_{k=0}^{K_\lambda}\frac{\log^k(\eta)}{k!} \sum_{n=0}^{\infty}{\mathbb
    H}_{\lambda,k,n}(i,j)\eta^n\,,
\end{equation}
where, for the coefficient matrices on the r.h.s., we used again subscripts to indicate the eigenvalue, the logarithmic power and the eta power, while $i$ and $j$ are matrix indices.

In the following, we denote by $\mathbb{H}(j)$ the solution on the $j$-th column of $\mathbb{H}$ and by $\lambda_j$ the eigenvalue on the $j$-th column of the Jordan matrix $\mathbb{A}_0$.
  
We then proceed as follows:

\paragraph{Step 1.}
  We initialize $\mathbb{H}_{\lambda_j,0,0}(i,j) = \delta_{i,j}$,
  while, for any other $\lambda \ne \lambda_j$, we set $\mathbb{H}_{\lambda,0,0}(i,j) = 0$.
  The latter implies, through Eqs.\,\eqref{eq:rec-log} and~\eqref{eq:rec-eta}, $\mathbb{H}_{\lambda,k,n}(j) = 0$ $ \forall \,k, n$ if $\lambda \ne \lambda_j$, so that the solution $\mathbb{H}(j)$ only exhibits the single eigenvalue $\lambda_j$.
  It is well known from standard textbooks (see e.g. \cite{wasow1965asymptotic}) that such initialization leads to the general solution of the homogeneous equation, so that every solution to Eq.\,\eqref{eq:de-block} can be expressed as linear combination of the columns $\mathbb{H}(j)$.

\paragraph{Step 2.}
  We use Eq.\,\eqref{eq:rec-log} with $n=0$,
  \begin{equation}
    \label{eq:rec-hom-log}
    {\mathbb H}_{\lambda_j,k+1,0} (j) = \frac{1}{g_0}\left( {\mathbb B}_0-\lambda_j g_0 \mathbbm{1} \right) {\mathbb H}_{\lambda_j,k,0}(j) = (\mathbb{A}_0 - \lambda_j \mathbbm{1}) {\mathbb H}_{\lambda_j,k,0}(j) \,,
  \end{equation}
  to fill the $\eta^0$ coefficients of the Taylor series that multiplies every logarithmic power.

  The matrix $\mathbb{A}_0$ is in Jordan form.
  Each index $j$ lies within a Jordan chain $a(j)$ of length $L_{a(j)}$ and eigenvalue $\lambda_j$.
  With each chain one can associate a vector subspace, which is the span of the vectors in the chain.
  This subspace is invariant under the action of the Jordan matrix.
  Since $(\mathbb{A}_0 - \lambda_j \mathbbm{1})$ is a Jordan matrix too (every eigenvalue being just shifted by $-\lambda_j$), Eq.\,\eqref{eq:rec-hom-log} can generate non-zero components at higher logarithmic powers only within the Jordan subspace of the chain $a(j)$.
  
  Note that Eq.\,\eqref{eq:rec-hom-log} can relate a given logarithmic order directly to the lowest one:
  \begin{align*}
    {\mathbb H}_{\lambda_j,k,0} (j) &= (\mathbb{A}_0 - \lambda_j \mathbbm{1}) {\mathbb H}_{\lambda_j,k-1,0}(j) \\
    &= (\mathbb{A}_0 - \lambda_j \mathbbm{1})^2 {\mathbb H}_{\lambda_j,k-2,0}(j) \\
    &= \ldots \\
    &= (\mathbb{A}_0 - \lambda_j \mathbbm{1})^k {\mathbb H}_{\lambda_j,0,0}(j) \,. \numthis
  \end{align*}

  Since $\lambda_j$ is subtracted from the diagonal of $\mathbb{A}_0$, the corresponding Jordan block $\mathbb{J}_{a(j)}$ in $(\mathbb{A}_0 - \lambda_j \mathbbm{1})$
  takes the form
  \begin{equation}
    \mathbb{J}_{a(j)} =
    \begin{pNiceMatrix}[columns-width=auto]
      0 & 1 & 0 & \dots & 0 \\
      0 & 0 & 1 & \dots & 0 \\
      \vdots & \vdots & \vdots & \ddots & \vdots \\
      0 & 0 & 0 & \ldots & 1 \\
      0 & 0 & 0 & \ldots & 0
    \end{pNiceMatrix}
    \,,
  \end{equation}
  which is nilpotent with degree $L_{a(j)}$:
  \begin{equation}
    \mathbb{J}_{a(j)}^{L_{a(j)}} = 0 \,.
  \end{equation}
  This implies that, at order zero in $\eta$, there are no logarithmic powers higher than $L_{a(j)}-1$ among the solutions $\mathbb{H}(j)$ associated with the Jordan chain $a(j)$.
  In particular, at logarithmic order zero the $m$-th vector of the chain starts with $1$ as the $m$-th component along the subspace; then, this component is shifted one position up after every application of $\mathbb{J}_{a(j)}$ and disappears after $m$ iterations:
  \begin{equation}
    \left(
    \begin{array}{c}
      0 \\ \vdots \\ 0 \\ 1 \\ 0 \\ \vdots
    \end{array}
    \right)
    \rightarrow
    \left(
    \begin{array}{c}
      0 \\ \vdots \\ 1 \\ 0 \\ 0 \\ \vdots
    \end{array}
    \right)
    \rightarrow \ldots \rightarrow
    \left(
    \begin{array}{c}
      1 \\ \vdots \\ 0 \\ 0 \\ 0 \\ \vdots
    \end{array}
    \right)
    \rightarrow
    \left(
    \begin{array}{c}
      0 \\ \vdots \\ 0 \\ 0 \\ 0 \\ \vdots
    \end{array}
    \right) \,.
  \end{equation}
  Therefore, the $m$-th vector of the chain receives $m-1$ logarithmic powers at order zero in $\eta$.

\paragraph{Step 3.}
  At this point we have computed every non-zero coefficient solution $\mathbb{H}_{\lambda_j, k, 0}(j)$.
  We now use Eq.\,\eqref{eq:rec-eta} to fill the coefficients of the higher $\eta$ powers for each logarithmic order.

  Note that even if, for some $k$, $\mathbb{H}_{\lambda_j, k, 0}(j) = 0$ within its Jordan subspace, the term ${\mathbb B}_1 {\mathbb H}_{\lambda_j,k,1}$ in  Eq.\,\eqref{eq:rec-eta} might pick up non-zero contributions from other subspaces.
  Only if all the components of $\mathbb{H}_{\lambda_j, k, 0}(j)$ are vanishing we can be sure that the Taylor series multiplying the logarithmic power $k$ is zero.
  Therefore, the maximum logarithmic power $K_{\lambda_j}$ appearing in the solution $\mathbb{H}(j)$ is related to the length of the longest Jordan chain with eigenvalue $\lambda_j$:
  \begin{equation}
    K_{\lambda_j} = \max\{L_{a(j')}: \lambda_{j'} = \lambda_j\} - 1 \,.
  \end{equation}

  With this said, we can compute $\mathbb{H}_{\lambda_j, K_{\lambda_j}, 1}(j)$, $\mathbb{H}_{\lambda_j, K_{\lambda_j}, 2}(j)$,~$\dots$ for the logarithmic order $k=K_{\lambda_j}$ using Eq.\,\eqref{eq:rec-eta} with $\mathbb{H}_{\lambda,K_{\lambda_j}+1,n-i}(j) = 0$.
  Then, the same equation can be used to fill the coefficients of the Taylor series that multiply the logarithmic powers $k=K_{\lambda_j}-1$, $K_{\lambda_j}-2$, \dots, $0$.


\subsubsection{Particular solution of the non-homogeneous equation}
We look for a particular solution $\textbf{P}(\eta)$ of Eq.\,\eqref{eq:de-block} in the form of Eq.\,\eqref{eq:sol}.
From the recurrence Eqs.\,\eqref{eq:rec-log} and~\eqref{eq:rec-eta} we see that all the eigenvalues of the non-homogeneous term $\mathbf{Y}(\eta)$ contribute to the solution.
On the other hand, block eigenvalues that are not in $\textbf{Y}(\eta)$ (i.e. they
do not show up in the solution for the previous blocks)
are already accounted for in the solution of the homogeneous equation.
Therefore, the particular solution only includes the eigenvalues of the non-homogeneous term.

To build the particular solution, we distinguish between the eigenvalues present only in $\mathbf{Y}(\eta)$ and those shared by $\mathbf{Y}(\eta)$ and the block.
In the following, we denote by $K_\lambda[Y]$ the maximum logarithmic power appearing in the $\lambda$-contribution to the non-homogeneous term and, for block eigenvalues, we indicate by $K_\lambda[A]$ the length of the longest Jordan chain with eigenvalue $\lambda$.

\paragraph{Eigenvalues only present in the non-homogeneous term.}
If a given eigenvalue $\lambda$ in $\mathbf{Y}(\eta)$ is not a block eigenvalue, the matrix $(\mathbb{B}_0 - (\lambda + n) \mathbbm{1})$ in Eq.\,\eqref{eq:rec-eta} is invertible also for $n=0$.
Therefore, we can use Eq.\,\eqref{eq:rec-eta} to build the all the Taylor coefficients, including the constant term.

We start at the logarithmic order $k = K_{\lambda,Y}$, setting $\mathbf{P}_{\lambda,K_{\lambda,Y}+1,n-i} = 0$ in Eq.\,\eqref{eq:rec-eta} with $n=0$:
\begin{equation}
  {\mathbf P}_{\lambda,K_\lambda[Y],0} = - g_0^{-1} \left( {\mathbb A}_0 - \lambda \mathbbm{1} \right)^{-1} \mathbf{Y}_{\lambda,K,0} \,.
\end{equation}
The maximum logarithmic power is indeed $K_{\lambda}[Y]$ as we can see plugging this result in Eq.\,\eqref{eq:rec-log} to check the next logarithmic order:
\begin{align}
  {\mathbf P}_{\lambda,K_{\lambda}[Y]+1,0} &= \left( {\mathbb A}_0 - \lambda \mathbbm{1} \right) {\mathbf P}_{\lambda,K_{\lambda}[Y],0} + g_0^{-1} {\mathbf Y}_{\lambda,K_{\lambda}[Y],0} \\
                        &= - g_0^{-1} \left( {\mathbb A}_0 - \lambda \mathbbm{1} \right) \left( {\mathbb A}_0 - \lambda \mathbbm{1} \right)^{-1} \mathbf{Y}_{\lambda,K_{\lambda}[Y],0} + g_0^{-1} {\mathbf Y}_{\lambda,K_{\lambda}[Y],0} \\
                        &= - g_0^{-1} \mathbf{Y}_{\lambda,K_{\lambda}[Y],0} +
                          g_0^{-1} {\mathbf Y}_{\lambda,K_{\lambda}[Y],0} = 0 \,.
\end{align}

The remaining coefficients for $k=K_{\lambda}[Y]$ can be computed using Eq.\,\eqref{eq:rec-eta} with $n>0$.
Then, one can proceed by using the same equation for the orders $k=K_{\lambda}[Y]-1$, $K_{\lambda}[Y]-2$, \dots, $0$ starting from $n=0$ up to the desired power of $\eta$.

\paragraph{Eigenvalues shared by the non-homogeneous term and the block}
For these eigenvalues $\lambda$ we start by initializing $\mathbf{P}_{\lambda,0,0} = 0$ and using Eq.\,\eqref{eq:rec-log} with $n=0$ to compute the constant term that multiplies every logarithmic power up to $k = K_{\lambda}[Y] + 1$:
\begin{equation}
  {\mathbf P}_{\lambda,k+1,0} = \left( {\mathbb A}_0 - \lambda \mathbbm{1} \right) {\mathbf P}_{\lambda,k,0} + g_0^{-1} {\mathbf Y}_{\lambda,k,0} \,, \quad k \le K_\lambda[Y]\,.
\end{equation}
When proceeding with $k > K_{\lambda}[Y] + 1$, the non-homogeneous term does not contribute anymore:
\begin{align*}
  \label{eq:par-log}
  {\mathbf P}_{\lambda,K_{\lambda}[Y]+1+k',0} &= \left( {\mathbb A}_0 - \lambda \mathbbm{1} \right) {\mathbf P}_{\lambda,K_{\lambda}[Y]+k',0} \\*
    &= (\mathbb{A}_0 - \lambda \mathbbm{1})^2 {\mathbf P}_{\lambda,K_{\lambda}[Y]+k'-1,0} \\*
    &= \ldots \\*
    &= (\mathbb{A}_0 - \lambda \mathbbm{1})^{k'} {\mathbf P}_{\lambda,K_{\lambda}[Y]+1,0} \,, \quad k' \ge 1 \,. \numthis
\end{align*}

For the vector components along the Jordan sub-spaces of
eigenvalue $\lambda$, the situation is similar to what happens for the homogeneous equation: the blocks in the matrix $(\mathbb{A}_0 - \lambda \mathbbm{1})$ are nilpotent, so that logarithmic powers greater than $K_{\lambda} = K_{\lambda}[Y]+K_{\lambda}[A]$ are zero.
On the other hand, the blocks in $(\mathbb{A}_0 - \lambda \mathbbm{1})$ associated with other Jordan chains are not nilpotent and Eq.\,\eqref{eq:par-log} can fill the corresponding vector components up to any logarithmic order with no limit.
Moreover, these orders cannot be canceled by a solution of the homogeneous equation, since the latter only has a finite number of logarithmic powers.

We note that, if $\mathbf{P}'(\eta)$ verifies Eq.\,\eqref{eq:rec-log} with $\mathbf{Y}(\eta) = 0$, $\mathbf{P}(\eta) + \mathbf{P}'(\eta)$ is still a solution for the same equation with $\mathbf{Y}(\eta) \ne 0$.
Therefore, in order to find a particular solution with limited logarithmic powers along all vector components, we look for an auxiliary term $\mathbf{P}'(\eta)$ that cancels the exceeding logarithmic powers.
Since we have to cancel these powers along multiple vector components, we build more than one auxiliary term and arrange them as columns of a matrix ${\mathbb P}'$ in the form
\begin{equation}
  \mathbb{P}'(i,j) = \sum_{\lambda\in S[A]} \mathbb{P}'_{\lambda}(i,j) = \sum_{\lambda\in S[A]} \eta^\lambda\, \mathbb{P}'_{\lambda,0,0}(i,j) \,,
\end{equation}
where $S[A]$ is the set of block eigenvalues and $\mathbb{P}'_{\lambda}$ is the contribution of a single $\lambda$.
We initialize with
\begin{equation}
  \mathbb{P}'_{\lambda, 0, 0}(i, j) =
  \begin{cases}
    \delta_{i, j} & \lambda \ne \lambda_j \\
    0 & \lambda = \lambda_j \,,
  \end{cases}
\end{equation}
so that, for each column $j$, all eigenvalues but the column eigenvalue $\lambda_j$ contribute and they do so with non-zero components outside the
Jordan sub-spaces associated with $\lambda_j$.
Using Eq.\,\eqref{eq:rec-log} with $n=0$ we can fill the constant terms for the logarithmic powers up to $k = K_{\lambda} + 1 = K_{\lambda}[Y] + K_{\lambda}[A] + 1$ since, once again, the corresponding blocks in $(\mathbb{A}_0 - \lambda \mathbbm{1})$ are not nilpotent.

Once both $\mathbf{P}$ and $\mathbb{P}'$ are filled up to $k = K_{\lambda}+1$, we build a linear combination of column solutions
\begin{equation}
  \mathbf{P}' = \sum_{\lambda \in S[A], j} c'_{\lambda,j} \mathbb{P}'_{\lambda}(j)
\end{equation}
imposing the cancellation of the logarithmic power $k = K_{\lambda} + 1$:
\begin{align*}
  \label{eq:log-match}
  0 &= \mathbf{P}_{\lambda, K_{\lambda} + 1, 0} + \mathbf{P}'_{\lambda, K_{\lambda}+1, 0} = \mathbf{P}_{\lambda, K_{\lambda} + 1, 0} + \sum_{j} c'_{\lambda,j} \mathbb{P}'_{\lambda, K_{\lambda} + 1, 0}(j) \numthis \,.
\end{align*}
Selecting from Eq.\,\eqref{eq:log-match} the vector components outside the Jordan sub-spaces of $\lambda$, we obtain a linear system of equations for the unknown coefficients $c'_{\lambda,j}$.
Once the system is solved, the constant term that multiplies every logarithmic power is redefined as
\begin{equation}
  \mathbf{P}_{\lambda, k, 0} \rightarrow \mathbf{P}_{\lambda, k, 0} + \mathbf{P}'_{\lambda, k, 0} \,, \quad k = 0
  \,.
\end{equation}
Here, $\mathbf{P}_{\lambda, K_{\lambda}+1, 0} = 0$ by construction and Eq.\,\eqref{eq:rec-log} with $n=0$ shows us that the same is true for any other power $k > K_{\lambda}+1$:
\begin{equation}
  {\mathbf P}_{\lambda,K_{\lambda}+1+k',0} = (\mathbb{A}_0 - \lambda \mathbbm{1})^{k'} {\mathbf P}_{\lambda,K_{\lambda}+1,0} = 0 \,, \quad k' \ge 1 \,.
\end{equation}
These logarithmic orders can be set to zero also at higher $\eta$-orders,
\begin{equation}
  \mathbf{P}_{\lambda,k,n} = 0 \quad \forall\,\, k > K_{\lambda} \,, n > 0 \,,
\end{equation}
since this is consistent with Eq.\,\eqref{eq:rec-eta}.

Finally, we can proceed as usual by using Eq.\,\eqref{eq:rec-eta} with $n = 0$, $1$, \dots, to fill any order of the Taylor series that multiply the logarithmic orders $k = K_{\lambda}$, $K_{\lambda}-1$, \dots, $0$, thus completing the particular solution with any necessary non-zero coefficient.

\subsubsection{Matching with boundary conditions}
\label{sec:match-with-bound}
With a particular solution $\widetilde{\mathbf{P}}(\eta)$ of Eq.\,\eqref{eq:de-block} and $L$ independent solutions of its associated homogeneous equation arranged as columns of $\widetilde{\mathbb{H}}(\eta)$, we can write down the general solution as
\begin{equation}
  \widetilde{\mathbf{I}}(\eta) = \widetilde{\mathbf{P}}(\eta) + \sum_{j = 1}^L c_j\, \widetilde{\mathbb{H}}(j; \eta) \,.
\end{equation}

The $L$ unknown coefficients $c_j$ associated with the columns $\widetilde{\mathbb{H}}(j)$ can be determined imposing that the solution evaluated in the boundary point $\eta = \eta_{bc}$ matches the transformed boundary $\widetilde{\mathbf{I}}_{bc}$:
\begin{equation}
  \widetilde{\mathbf{P}}(\eta_{bc}) + \sum_{j = 1}^L c_j\, \widetilde{\mathbb{H}}(j; \eta_{bc}) = \widetilde{\mathbf{I}}_{bc} \,.
\end{equation}
This results in a linear system of $L$ equations with matrix $\widetilde{\mathbb{H}}(\eta_{bc})$ and constant term given by $\widetilde{\mathbf{I}}_{bc} - \widetilde{\mathbf{P}}(\eta_{bc})$.
When evaluating this quantities, analytic continuation is needed for the complex logarithm $\Log(\eta_{bc})$ (see section~\ref{sec:analyt-cont}).

The solution coefficients $c_j$ completely determine $\widetilde{\mathbf{I}}(\eta)$ that can be transformed back to the original non-Fuchsian basis,
\begin{equation}
  \mathbf{I}(\eta) = \mathbb{T}(\eta)\, \widetilde{\mathbf{I}}(\eta) \,,
\end{equation}
and evaluated in any other point within the radius of convergence of the series.

\subsection{Analytic continuation}\label{sec:analyt-cont}
As discussed in Section~\ref{sec:DEs-for-MIs}, when a singular point is met along the phase-space line, we cross it by solving the DEs around it.
As we can see from the ansatz in Eq.\,\eqref{eq:sol}, the solution is expressed in terms of complex logarithms that exhibit a \textit{branch cut} in the $\eta$-complex plane originating from the singular point.
In principle, the orientation of the branch cut can be arbitrarily chosen, selecting different definitions of the logarithm function appearing in the ansatz.
By doing so, one can find many solutions to the DEs matching the same boundary conditions.
If the singular point is not a branch point for the MIs, then the orientation of the branch cut in the $\eta$-complex plane is irrelevant, since different choices lead to the same result when evaluating the solution in the next point of the path.
On the other hand, if the MIs do have a branch point in the singular point, the choice of the branch cut affects the solution.
In fact, the branch cut is mapped by the parameterization in Eq.\,\eqref{eq:line} from the $\eta$-complex plane to the phase-space of the kinematic invariants, where the MIs have a branch cut with a specific orientation.
Therefore, the branch cut in the $\eta$-plane must be selected consistently to ensure it maps correctly onto the branch cut of the MIs.


The branch points of a FI are associated with its \textit{Cutkosky cuts}, defined as any deletion of a set of edges from the corresponding Feynman diagram resulting into two disjoint diagrams.
Let $\alpha_1 t_1 + \alpha_2 t_2 + \dots + \alpha_N t_N$ be the squared momentum flowing through the cut expressed as a linear combination with coefficients $\alpha_i$ of $N$ external kinematic invariants $t_i$;
also, let $m_1, m_2, \dots, m_M$ be the masses of the $M$ cut propagators; we can define the \textit{Cutkosky invariant} $z$ associated with the cut as
\begin{equation}
  \label{def:cutk-inv}
  z \equiv \alpha_1 t_1 + \alpha_2 t_2 + \dots + \alpha_N t_N - (m_1+ m_2 + \dots + m_M)^2 \,.
\end{equation}
The FI has a branch cut in the $z$-complex plane defined by $z\ge0$, originating from a branch point in $z=0$.
According to the Feynman prescription, the value of the FI on the branch cut is obtained approaching the positive real axis from above.
These conditions can be transferred from the $z$- to the $\eta$-plane using the map $z(\eta)$ defined by plugging the parameterization of Eq.\,\eqref{eq:line} into Eq.\,\eqref{def:cutk-inv}:
{
\allowdisplaybreaks[0]
\begin{align}
  \label{def:branch-point}
  z(\eta) &= 0 \qquad \text{branch point} \,, \\
  \label{def:branch-cut}
  z(\eta) &\ge 0 \qquad \text{branch cut} \,.
\end{align}
}

In the following, we use these equations to find branch points and branch cuts in the $\eta$-complex plane, starting from the simple case where the cut masses are fixed along the path and then extending to the general case of varying masses.

Before proceeding, it is useful to establish some notation that is employed throughout the rest of this section.
First of all, we use $\log(x)$ to indicate the real-valued logarithm function and $\Log(x)$ for its complex-valued continuation.
We measure angles in the complex plane counter-clockwise
so that $\varphi=0$ identifies the positive real axis, while $\varphi > 0$ $(\varphi < 0)$ represents the upper (lower) half-plane.
Also, we consider $\arg(x) \in [ -\pi, \pi[$ and, for any angle $\varphi$, we introduce the function $\arg^{(\varphi)}(x) \in [\varphi, \varphi + 2\pi[$ that picks up angles counter-clockwise starting from $\varphi$.

When a branch cut is in the lower half-plane at an angle $\varphi \in [ -\pi, 0[$, we assign the imaginary part of $\Log(x)$ within $[\varphi, \varphi + 2 \pi[$, while for a branch cut in the upper half-plane at angle $\varphi \in [ 0, \pi[$ we shift down by $2 \pi$ and take the imaginary part in $[\varphi - 2 \pi, \varphi[$.
With this choice, the logarithm of a positive real number is always real, while for negative numbers the imaginary part of the logarithm is $\pi$  when the branch cut is in the lower half-plane, $-\pi$ when it is in the upper one.

\subsubsection{Cut with fixed masses}
\label{sec:fixed-masses}
When the cut masses are kept constant, only the external invariants depend on $\eta$ and we have
\begin{equation}
  \left\{
  \begin{array}{ccl}
    t_1(\eta) &=& t_{1,i} + \eta \left( t_{1,f} - t_{1,i} \right) \\
    \vdots & & \\
    t_N(\eta) &=& t_{N,i} + \eta \left( t_{N,f} - t_{N,i} \right)
  \end{array}
  \right.
\end{equation}
which, inserted into Eq.\,\eqref{def:cutk-inv}, gives
\begin{align*}
  z(\eta) &= \sum_{k=1}^N \alpha_k t_{k,i} - \left( \sum_{k=1}^M m_k \right)^2 + \eta \sum_{k=1}^N \alpha_k \left( t_{k,f} - t_{k,i} \right) \\
  &\equiv z_i + \eta \left( z_f - z_i \right) \numthis  \,,
\end{align*}
where
\begin{equation}
  z_i \equiv \sum_{k=1}^N \alpha_k t_{k,i} - \left( \sum_{k=1}^M m_k \right)^2, \qquad z_f \equiv \sum_{k=1}^N \alpha_k t_{k,f} - \left( \sum_{k=1}^M m_k \right)^2 \,,\\
\end{equation}
while the difference $z_f - z_i$ does not depend on the masses:
\begin{equation}
  z_f - z_i = \sum_{k=1}^N \alpha_k \left( t_{k,f} - t_{k,i} \right) \,.
\end{equation}

The branch point $\eta_b$ can be obtained from $z(\eta_b) = 0$, that is,
\begin{equation}
  \label{eq:branch-point}
  \eta_b = - \frac{z_i}{z_f - z_i} \,.
\end{equation}
We then look for a branch cut parameterized as an half line originating from $\eta_b$ at angle $\varphi_b$ w.r.t. the positive real axis, i.e. $\left\{ \eta = \eta_b + r e^{i \varphi_b}, r \ge 0 \right\}$.
To this purpose, we use Eq.\,\eqref{def:branch-cut} whose l.h.s. is
\begin{align*}
  \label{stp:find-branch-cut}
  z ( \eta_b + r e^{i \varphi_b} ) &= z_i + \left( \eta_b + r e^{i \varphi_b} \right) \left( z_f - z_i \right) \\
    &= z_i + \eta_b \left( z_f - z_i \right) + r e^{i \varphi_b} \left( z_f - z_i \right) \\
    &= r e^{i \varphi_b} \left( z_f - z_i \right) \numthis \,,
\end{align*}
where the defining equation for $\eta_b$, $z_i + \eta_b \left( z_f - z_i \right) = 0$, has been employed.
In order for the r.h.s. of Eq.\,\eqref{stp:find-branch-cut} to be real and positive, we must have
\begin{equation}
  \label{eq:branch-cut-angle}
  \varphi_b = \arg \left( z_i - z_f \right) \,,
\end{equation}
which, together with Eq.\,\eqref{eq:branch-point}, completely specifies the location of the branch cut.

Now that we know where the branch cut is, we can proceed with the evaluation of any logarithm centered in $\eta_b$.
Recall that we are interested in analyzing a branch point lying on the path, therefore $\eta_b \in ]0, 1[$ for our use case
.
Also, we need the value of the complex logarithm in its left and right nearest neighbors on the path, which also belong to $]0, 1[$ and thus are real.

In case the branch cut does not lie on the real axis, we have
\begin{itemize}
\item
  $ \arg \left( z_i - z_f \right) \in ]-\pi, 0[$ $\implies $ branch cut in the lower half-plane:
  \begin{equation}
    \Log(\eta - \eta_b) =
    \begin{cases}
      \log(\eta - \eta_b) & \eta > \eta_b \\
      \log(\eta_b - \eta) + i \pi & \eta < \eta_b
    \end{cases}
  \end{equation}

\item
  $ \arg \left( z_i - z_f \right) \in ]0, \pi[$ $\implies $ branch cut in the upper half-plane:
  \begin{equation}
    \Log(\eta - \eta_b) =
    \begin{cases}
      \log(\eta - \eta_b) & \eta > \eta_b \\
      \log(\eta_b - \eta) - i \pi & \eta < \eta_b
    \end{cases}
  \end{equation}

\end{itemize}

If $z_i - z_f$ is real, the branch cut is on the real axis and we need to evaluate the complex logarithm on its cut.
In this case, the result depends on whether the real axis of $z$ must be approached from above, as in the Feynman prescription, or below.
We can parameterize these two limits as a clockwise or counter-clockwise rotation in the $z$-complex plane, respectively.
To approach the axis from above (below), we consider $z = a e^{i \varepsilon}$ in the limit $\varepsilon \rightarrow 0^+$ ($\varepsilon \rightarrow 0^-$) for some $a>0$.
The angle $\varepsilon$ corresponds, through the map $z(\eta)$, to an angle $\delta$ w.r.t. the branch cut in the $\eta$-plane.
The sign of $\delta$ tells us if the complex logarithm must be analytically continued on its branch cut through a clockwise or counter-clockwise rotation in $\eta$.
We have
{
\allowdisplaybreaks[0]
\begin{align*}
  a e^{i \varepsilon} &= z_i + ( \eta_b + r e^{i (\varphi_b + \delta)} ) (z_f - z_i) \\
  &= r e^{i (\varphi_b + \delta)} (z_f - z_i) \,,
\end{align*}
}
meaning that
\begin{equation}
  \varepsilon = \varphi_b + \delta - \arg (z_i - z_f) = \delta \,.
\end{equation}

We then conclude that, if the positive real axis of $z$ is approached from above (below), the logarithm in $\eta$ must be analytically continued on its branch-cut with a clockwise (counter-clockwise) rotation.
Therefore:

\begin{itemize}
\item
  $ z_i - z_f > 0 $ $\implies $ branch cut on the positive real axis approached from above (below):
  \begin{equation}
    \Log(\eta - \eta_b) =
    \begin{cases}
      \log(\eta - \eta_b) + i 2 \pi \theta_H(-\varepsilon) & \eta > \eta_b \\
      \log(\eta - \eta_b) + i \pi & \eta < \eta_b
    \end{cases}
  \end{equation}

\item
  $ z_i - z_f < 0 $ $\implies $ branch cut on the negative real axis approached from below (above):
  \begin{equation}
    \Log(\eta - \eta_b) =
    \begin{cases}
      \log(\eta - \eta_b) & \eta > \eta_b \\
      \log(\eta - \eta_b) - i \pi + i 2 \pi \theta_H(-\varepsilon) & \eta < \eta_b
    \end{cases}
  \end{equation}

\end{itemize}
The Feynman prescription is obtained for $\varepsilon>0$, so that the Heaviside step function $\theta_H(-\varepsilon)$ does not contribute.

We stress that, in general, the values of the imaginary part of the logarithm in $\eta > \eta_b$ and $\eta < \eta_b$ do not matter as long as their \textit{relative} value is the correct one.
In other words, we are allowed to add any integer multiple of $2 \pi$ to both of them (thus moving the logarithm on another Riemann sheet) as long as the location of the branch cut stays the same, since the coefficients for matching the boundary conditions in the previous path point account for the change in value.

\subsubsection{Cut with varying masses}
\label{sec:variable-masses}
Analytic continuation gets more involved when we vary also the cut masses while moving through the phase-space:
\begin{equation}
  \left\{
      \begin{array}{ccl}
        m^2_1 (\eta) & = & m^2_{1, i} + \eta (m^2_{1, f} - m^2_{1, i})\\
        \vdots &  & \\
        m^2_M (\eta) & = & m^2_{M, i} + \eta (m^2_{M, f} - m^2_{M, i}) \,.
      \end{array}
  \right.
\end{equation}
In fact, this time Eq.\,\eqref{def:cutk-inv} gives
\begin{equation}
  z(\eta) = \sum_{k=1}^N \alpha_k t_{k,i} + \eta \sum_{k=1}^N \alpha_k \left( t_{k,f} - t_{k,i} \right) - \left( \sum_{k=1}^M \sqrt{m^2_{k, i} + \eta (m^2_{k, f} - m^2_{k, i})} \right)^2
\end{equation}
and the square roots of the mass term on the r.h.s. might lead to a complicated branch cut in the $\eta$-complex plane.

For this reason in \Line, when a branch point has to be crossed, the \textit{linear} masses are varied instead of the squared ones:
\begin{equation}
  \left\{
      \begin{array}{ccl}
        m_1 (t) & = & m_{1, i} + \eta (m_{1, f} - m_{1, i})\\
        \vdots &  & \\
        m_M (t) & = & m_{M, i} + \eta (m_{M, f} - m^2_{M, i}) \,.
      \end{array}
  \right.
\end{equation}
This leads to the Cutkosky invariant
\begin{equation}
  \label{eq:cutk-inv-quad}
  z(\eta) = \sum_{k=1}^N \alpha_k t_{k,i} + \eta \sum_{k=1}^N \alpha_k \left( t_{k,f} - t_{k,i} \right) - \left[ \sum_{k=1}^M m_{k, i} + \eta \sum_{k=1}^M (m_{k, f} - m_{k, i}) \right]^2 \,,
\end{equation}
which is a quadratic polynomial. Its roots correspond to poles of the DEs, each serving as a branch point that generates a distinct branch cut.
However, we only solve the equations around one pole at a time, say $\eta_b \in ] 0, 1 [$, so we can focus on what happens within half the radius of convergence $\mathcal{R}(\eta_b)$ of the solution around this single branch point.

Due to the quadratic nature of the map $z(\eta)$, the branch point developing from $\eta_b$ is \textit{nonlinear}.
In principle, this does not represent a problem since, when choosing a branch for the multi-valued complex logarithm $\Log(x)$, we are not forced to use a straight line to define the boundary between two Riemann sheets.
Consider for example any \textit{curved line} $\gamma : [0, + \infty [ \rightarrow \mathbb{C}$ starting from the origin and stretching out to infinity without coiling, i.e. for all $r > 0$ there is only one point of $\gamma$ at distance $r$ from the origin.
Under this assumptions we can find a line parameter $r$ such that $\abs{\gamma(r)} = r$ for all $r\ge0$.
We can then define the logarithm with a curved branch cut $\gamma$ as the function that, at any distance $\abs{x}=r$ from the origin, is equal to the logarithm with a (linear) branch cut at angle $\arg(\gamma(r))$:
\begin{equation}
  \Log(x) = \log(\abs{x}) + i \text{Arg}^{(\arg(\gamma(\abs{x})))} (x) \,.
\end{equation}
Branch cut represented by more complicated, coiling curves could also be defined; however, this is not necessary for our purposes.

In our use case, the branch cut is $\left\{ \eta = \eta_b + \gamma(r), r \ge 0 \right\}$ and we only need to evaluate the logarithm in two points $\eta_b \pm r_b$ that are on the real axis at a fixed distance $r_b \equiv \mathcal{R}(\eta_b)/2$ from the branch point $\eta_b$.
Therefore, all we need to know is whether $\gamma(r_b)$ is in the upper or lower half complex plane.
We can then reason as follows: $\gamma(r_b)$ is located, say, in the lower half-plane if and only if, starting from $\eta_b - r_b$, a point $P$ moving counter-clockwise around $\eta_b$ on the circle of radius $r_b$ meets the branch cut before getting to $\eta_b + r_b$.
This in turn happens if and only if the Cutkosky invariant $z(P)$ crosses the positive real axis in the $z$-plane.
Such a condition can be established comparing the angles of the final and initial points $z(\eta_b \pm r_b)$ since, as proven below, $z(P)$ also moves counter-clockwise in its complex plane with no changes in direction.
We have:
\begin{itemize}
\item
  $0 < \arg^{(0)}(z(\eta_b + r_b)) < \arg^{(0)}(z(\eta_b - r_b))$ $\implies$ branch cut in the lower half-plane: 
  \begin{equation}
    \Log(\eta - \eta_b) =
    \begin{cases}
      \log(\eta - \eta_b) & \eta = \eta_b + r_b \\
      \log(\eta - \eta_b) + i \pi & \eta = \eta_b - r_b
    \end{cases}
  \end{equation}

\item
    $0 < \arg^{(0)} (z(\eta_b - r_b)) < \arg^{(0)} (z(\eta_b + r_b))$ $\implies$ branch cut in the upper half-plane:
  \begin{equation}
    \Log(\eta - \eta_b) =
    \begin{cases}
      \log(\eta - \eta_b) & \eta = \eta_b + r_b \\
      \log(\eta - \eta_b) - i \pi  & \eta = \eta_b - r_b
    \end{cases}
  \end{equation}
\end{itemize}

Once again the edge cases are when $\gamma(r_b)$ is on the positive or negative real axis, which happens if the final or initial points $z(\eta_b \pm r_b)$ are on the positive real axis in the $z$-plane, respectively:
\begin{itemize}
\item
  $\arg^{(0)}(z(\eta_b + r_b)) = 0$ $\implies $ branch cut on the positive real axis:
  \begin{equation}
    \Log(\eta - \eta_b) =
    \begin{cases}
      \log(\eta - \eta_b) + i 2 \pi \theta_H(-\varepsilon) & \eta > \eta_b \\
      \log(\eta - \eta_b) + i \pi & \eta < \eta_b
    \end{cases}
  \end{equation}

\item
  $\arg^{(0)}(z(\eta_b - r_b)) = 0$ $\implies $ branch cut on the negative real axis:
  \begin{equation}
    \Log(\eta - \eta_b) =
    \begin{cases}
      \log(\eta - \eta_b) & \eta > \eta_b \\
      \log(\eta - \eta_b) - i \pi + i 2 \pi \theta_H(-\varepsilon) & \eta < \eta_b
    \end{cases}
  \end{equation}

\end{itemize}

We conclude the Section by showing that $z(P)$ moves counter-clockwise while $P$ circles around the branch point $\eta_b$ at distance $r_b$.
Let $\eta_b' \ne \eta_b$ be the other root of the quadratic polynomial $z(\eta)$, so that
\begin{equation}
  z(\eta) = z_2 (\eta - \eta_b) (\eta - \eta_b') \,,
\end{equation}
where $z_2$ is the coefficient of $\eta^2$, whose value can be easily read from Eq.\,\eqref{eq:cutk-inv-quad} but is not important for this proof.
The transformation $\eta = \eta_b + x (\eta_b - \eta_b')$ places the branch point $\eta_b$ at $x = 0$ and the other root $\eta_b'$ at $x=1$ in the complex plane of the newly introduced variable $x$, while changing $z$ to
\begin{equation}
  z = z_2 (\eta - \eta_b')^2 x (x - 1) \,.
\end{equation}

The complete rotation of a point P around $\eta_b$ at distance $r_b$ corresponds to $x = r e^{i \varphi}$, where $\varphi \in [0, 2 \pi]$ and $r_b = r \abs{P - \eta_b}$.
Since $\eta_b'$ is a pole of the DEs and we stay within half the distance between $\eta_b$ and its closest pole, we are sure that $r \le 1/2$.
In terms of $r$ and $\varphi$, the Cutkosky invariant becomes
\begin{equation}
  z = z_2 (\eta - \eta_b')^2 r e^{i \varphi} (r e^{i \varphi} - 1) \,.
\end{equation}
In order for $z$ to never change the direction of its rotation, the derivative of its argument must remain non-negative.
We have:
\begin{align*}
  \frac{d}{d \varphi} \arg^{(0)}(z) &= \frac{d}{d \varphi} \left[ \arg^{(0)}(z_2 (\eta - \eta_b')^2) + \arctan( \frac{r \sin (2 \varphi) - \sin(\varphi)}{r \cos(2 \varphi) - \cos(\varphi)}) \right] \\
  & = \frac{1 + 2 r^2 - 3 r \cos (\varphi)}{(\cos (\varphi) - r \cos (2 \varphi))^2} \,,
\end{align*}
which is non-negative when its numerator is, i.e. when $\cos (\varphi) \le (1 + 2 r^2) / 3 r$.
This holds true for sure if $(1 + 2 r^2) / 3 r \le 1$, which is the case for $r \le 1/2$ and $r \ge 1$.
Since $r \le 1/2$ in our case, we arrive at the desired conclusion.
Finally, note that if $1 / 2 < r < 1$ we would have $(1 + 2 r^2) / 3 r < 1$ and we could always find an angle $\varphi$ for which $\cos (\varphi) > (1 + 2 r^2) / 3 r$ and the derivative becomes negative.
This shows that staying within half the distance of the other pole $\eta_b'$ is indeed necessary
in order for this method to always succeed.

\subsection{Boundary conditions}
In order to compute the master integrals via the DE a boundary condition is needed.
There are well-established approaches to serve this purpose.

A first option is a direct numerical integration of $\mathbf{I}(\epsilon, \eta_{bc})$, that is usually carried out by Monte Carlo methods. In order to provide stable Monte Carlo integrations, UV and IR divergences have to be singled out from the integral. In addition, clever sampling methods are implemented and represent an active branch of research.
Tools that implement such techniques include  pySecDec~\cite{Borowka:2017idc} and FIESTA~\cite{Smirnov:2008py}.
The curse of dimensionality and the need to factorize divergences are common sources of slow convergence, which can negatively affect the accuracy of the Monte Carlo integration.

Analytical methods are extremely powerful, but unfortunately cannot be implemented straightforwardly. An analysis of the singularity structure of each MI have to be carried out.
Expansion by regions~\cite{Smirnov:1991jn,Beneke:1997zp,Heinrich:2021dbf} provides a systematic way for extracting boundary conditions and we are currently investigating its implementation in \Line{}.

In recent years, the auxiliary-mass flow method~\cite{Liu:2017jxz} has been introduced as a powerful approach for evaluating Feynman integrals in any phase-space point.
A Mathematica package implementing this method is publicly available~\cite{Liu:2022chg}.
Notably, the method can be used to determine boundary conditions for any fixed value of $\epsilon$, making it particularly well-suited for our purposes.
We have implemented this technique in \Line{} and in the following we present its operational details as applied within our framework.

\subsubsection{The AMFlow method}\label{sec:amf}
The basics of the method rely on the introduction of an auxiliary squared mass $\eta$ in all the denominators of the MIs. One can find the extended list $\boldsymbol{\mathcal{I}}({\epsilon, \bf s, \eta})$ of MIs for the new topology and the corresponding differential equations with respect to $\eta$ at fixed phase-space point $\bf s$:
\begin{equation}\label{eq:amfde}
  \frac{d}{d\eta}\boldsymbol{\mathcal{I}}(\epsilon, {\bf s, \eta}) = \mathbb{A}(\epsilon, \eta)\boldsymbol{\mathcal{I}}(\epsilon, {\bf s, \eta}) \,.
\end{equation}
These equations can be solved to propagate the MIs from infinity to zero along a path $\eta = \infty$, $\eta=\eta_1$, \dots, $\eta = \eta_{\text{last}}$, $\eta = 0$ in the lower half of the complex $\eta$-plane, recovering the physical values of the integrals computing the limit $\eta \rightarrow 0$.
The points $\eta = \infty$ and possibly $\eta = 0$ are the only regular singular ones, while the number of regular steps is affected by the singularities in the complex plane.

At infinity, the integrals admit the asymptotic expansion
\begin{equation}
  \mathcal{I}_i(\epsilon, {\bf s}, \eta) \widesim[2]{\eta \to \infty} \eta^{a_i}\left[\mathcal{I}_i(\epsilon, {\bf 0}, 1) + O(\eta) \right]\,, \qquad a_i = l\frac{d}{2} - \nu_i \,,
\end{equation}
where $\nu_i$ is the sum of the denominator exponents of $\mathcal{I}_i(\epsilon, {\bf s}, \eta)$ and $\mathcal{I}_i(\epsilon, {\bf 0}, 1)$ are unit-mass $l$-loop vacuum integrals. The latter are known in literature with their exact dependence on $\epsilon$ up to three-loop, and numerically up to five-loop.
In \Line{} we implemented the explicit expressions for the one- and two-loop vacuum integrals that we report below for convenience:
\begin{gather}
 \includegraphics[width=.09\textwidth,valign=c]{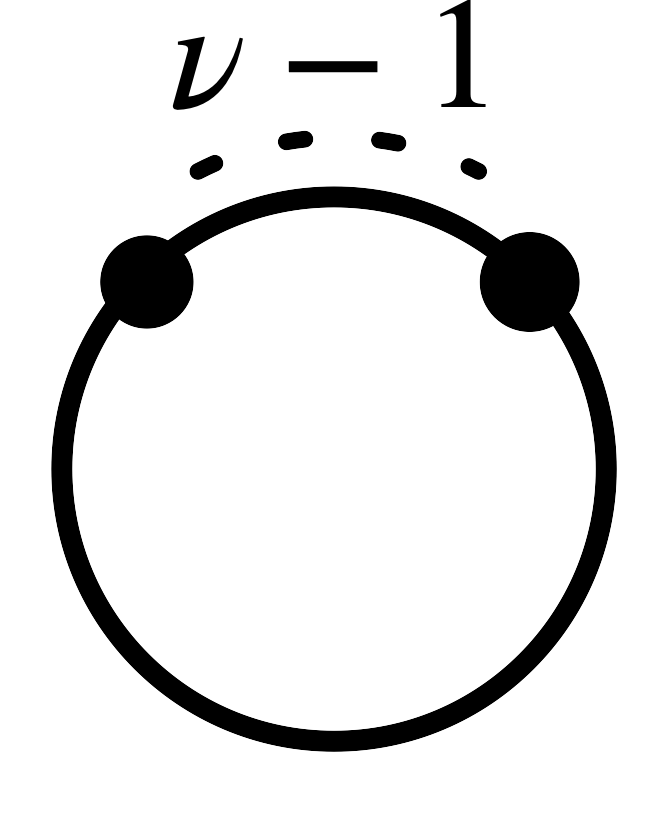} = (-1)^{\nu}\frac{\Gamma(\nu - 2 + \epsilon)}{\Gamma(\nu)} \,,\\
  \begin{aligned}
  \includegraphics[width=.1\textwidth,valign=c]{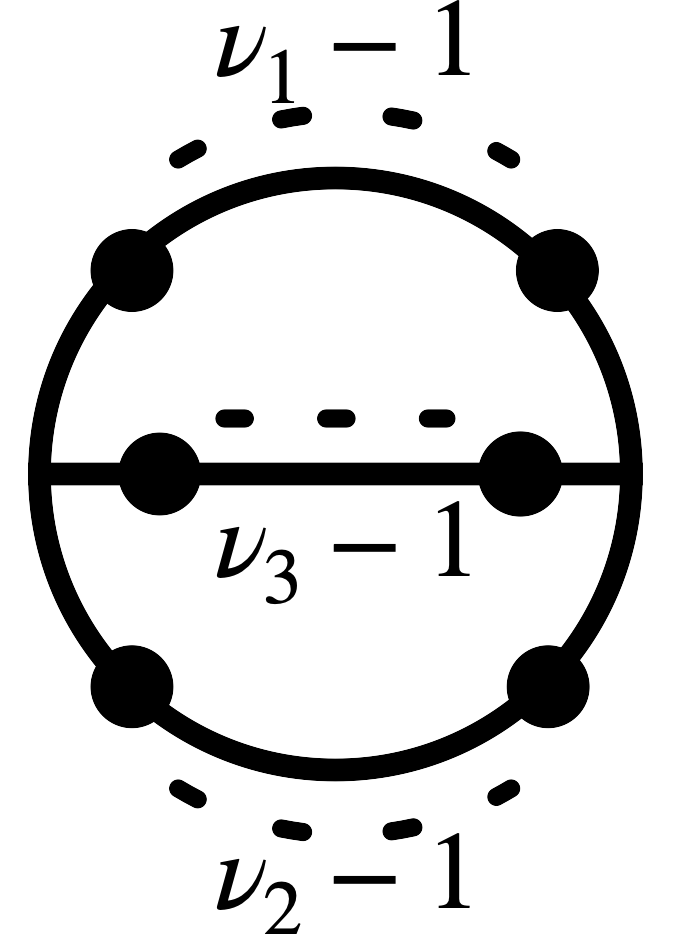} =& (-1)^{\nu} \biggr[
    \frac{\Gamma (\nu_3-2+\epsilon) \Gamma (\nu_1+\nu_2-2+\epsilon )}{\Gamma (\nu_3) \Gamma (\nu_1+\nu_2)} 
       \, _4F_3\left(
         \small\begin{matrix}
           2-\epsilon ,\nu_1,\nu_2,\nu_1+\nu_2-2+\epsilon \\
           \frac{\nu_1+\nu_2}{2},\frac{\nu_1+\nu_2}{2}+\frac{1}{2},3 -\nu_3-\epsilon
         \end{matrix} ;\frac{1}{4}\right) \\
    &+\frac{\Gamma (2-\nu_3-\epsilon) \Gamma (\nu_1+\nu_3-2+\epsilon) \Gamma (\nu_2+\nu_3-2+\epsilon) \Gamma (\nu+2 \epsilon-4)}{\Gamma (\nu_1) \Gamma (\nu_2) \Gamma (2-\epsilon ) \Gamma (\nu + \nu_3-4+2 \epsilon)}\\[1em]
    &\times \, _4F_3\left(
      \small\begin{matrix}
        \nu_3,\nu_1+\nu_3-2+\epsilon,\nu_2+\nu_3-2+\epsilon,\nu-4+2\epsilon\\
        \nu_3-1+\epsilon,\frac{\nu + \nu_3 - 4}{2}+\epsilon,\frac{\nu + \nu_3-3}{2}+\epsilon \end{matrix}; \frac{1}{4}\right)\biggr] \,,
     \end{aligned}
\end{gather}
where $\nu_i-1$ is the number of dots, and $\nu = \nu_1+\nu_2+\nu_3$.
These formulae can be used to compute the boundary conditions of Eq.\,\eqref{eq:amfde} for any one- or two-loop Feynman integrals.
For higher-loop calculations one needs the corresponding unit-mass vacuum integrals.
Alternatively, the approach where $\eta$ is only inserted in some of the propagators offers the possibility to compute boundary conditions by iteratively applying the AMFlow method~\cite{Liu:2021wks}.
The automated computation of boundary conditions for three- and higher-loop problems is beyond the aim of the present work.

The leading coefficients of the integrals at $|\eta|\to\infty$ are used as boundary condition when solving the $\eta$-DEs in Eq.\,\eqref{eq:amfde}.
One can change $\eta \rightarrow 1/\eta$ to center the problem around $\eta = 0$,
\begin{gather}
  \frac{d}{d\eta}\boldsymbol{\mathcal{I}}_{\infty}(\epsilon, \eta) = \mathbb{A}_{\infty}(\epsilon, \eta)\boldsymbol{\mathcal{I}}_{\infty}(\epsilon, \eta) \,,
\\ \boldsymbol{\mathcal{I}}_{\infty}(\epsilon, \eta) \equiv \boldsymbol{\mathcal{I}}\left(\epsilon, 1/\eta\right) \,,  \quad \mathbb{A}_{\infty}(\epsilon, \eta) \equiv -\frac{\mathbb{A}\left(\epsilon, 1/\eta\right)}{\eta^2} \numthis \,,
\end{gather}
and factor out the leading power behaviors $\eta^{-\mathbf{a}} \equiv (\eta^{-a_1}, \eta^{-a_2}, \dots)$ of the MIs,
\begin{gather}
  \frac{d}{d\eta}\widetilde{\boldsymbol{\mathcal{I}}}_{\infty}(\epsilon, \eta) = \widetilde{\mathbb{A}}_{\infty}(\epsilon, \eta) \widetilde{\boldsymbol{\mathcal{I}}}_{\infty}(\epsilon, \eta) \,,
  \\ \widetilde{\boldsymbol{\mathcal{I}}}_{\infty}(\epsilon, \eta) \equiv \eta^{-\mathbf{a}}\, \boldsymbol{\mathcal{I}}_{\infty}(\epsilon, \eta) \,,  \quad \widetilde{\mathbb{A}}_{\infty}(\epsilon, \eta) \equiv  \text{diag}(\eta^{\mathbf{a}}) \left(\mathbb{A}_{\infty}(\epsilon, \eta) + \frac{\text{diag}(\mathbf{a})}{\eta}\right) \text{diag}(\eta^{-\mathbf{a}}) \,,
\end{gather}
to solve for the Taylor series $\widetilde{\boldsymbol{\mathcal{I}}}_{\infty}(\epsilon, \eta)$ whose leading terms are the unit-mass vacuum integrals.


The solution is evaluated in the first regular point $\eta_1$ and used as boundary for the following propagations up to $\eta=\eta_{\text{last}}$.
Then, Eq.\,\eqref{eq:amfde} is solved around the (possibly) regular singular point $\eta=0$ matching the value of the MIs in $\eta = \eta_{\text{last}}$ and the final result is obtained computing the limit $\eta \rightarrow 0$ of the series solution.

\subsubsection{Boundary conditions through expansion by regions}
\label{sec:alternative-method}
The AMFlow method is based on expansion-by-regions in the infinite-mass limit, where the external kinematics is negligible w.r.t. the auxiliary mass $\eta$.
In fact, EBR constitutes a powerful tool that can be used to predict the leading behavior of FIs in kinematic configurations where an invariant becomes negligible w.r.t. the others.
When solving DEs around such points, one can obtain boundary conditions at all orders in $\epsilon$ by constraining the series solution to reproduce the behavior dictated by the EBR.

In Section~\ref{sec:examples}, we exploit this procedure for some examples where:
\begin{enumerate}
\item
  the EBR leading coefficients for the MIs of the simplest sub-topologies are known analytically;
\item
  the EBR leading coefficients for MIs of more complex sub-topologies can be determined from those of simpler sub-topologies by ensuring the solution exhibits the expected power behavior.
\end{enumerate}

The detailed analysis of such procedure in full generality goes beyond the aim of the present paper and will be the subject of future study.

\section{Examples}\label{sec:examples}
In the following, we evaluate at several phase-space points the coefficients of the $\epsilon$-expansion of the MIs for one- and two- loop topologies up to the $\epsilon^{2}$ and $\epsilon^{4}$ order, respectively.
The numerical results are presented with 16 digits of accuracy and arranged in tables.

We dub AMF$^0$ our in-house implementation of the AMFlow method that we use both for the computation of boundary conditions and to validate the propagations through the DEs with full kinematic dependence.
In some cases we make use of EBR to compute boundary conditions around poles of the DEs.

\subsection{One-loop triangle with six scales}
\label{sec:1L-triangle-full}
Let us consider the family of FIs associated with the full-scale one-loop triangle:
\begin{equation}
  \label{eq:1L-triangle-full}
  F^{\boldsymbol{\nu}} (\epsilon, s, p_1^2,p_2^2,m_1^2,m_2^2,m_3^2) = \int_k \frac{1}{D_1^{\nu_1} D_2^{\nu_2} D_3^{\nu_3}}
\end{equation}
with squared external momenta $s = (p_1+p_2)^2$, $p_1^2$, $p_2^2$ and inverse propagators
\begin{align*}
  D_1 &= k^2 - m_1^2 , \\
  D_2 &= (k + p_1)^2 - m_2^2 , \\
  D_3 &= (k + p_1 + p_2)^2 - m_3^2 . \numthis
\end{align*}
A basis of $7$ MIs for this topology is $I_i = F^{\boldsymbol{\nu}_i}$, with powers
{
\small
\begin{equation}
  \label{eq:1L-triangle-full-basis}
  \allowdisplaybreaks
  \begin{aligned}
    \boldsymbol{\nu}_1 &= (1,0,0) , &\quad
    \boldsymbol{\nu}_2 &= (0,1,0) , \\
    \boldsymbol{\nu}_3 &= (0,0,1) , &\quad
    \boldsymbol{\nu}_4 &= (1,1,0) , \\
    \boldsymbol{\nu}_5 &= (1,0,1) , &\quad
    \boldsymbol{\nu}_6 &= (0,1,1) , \\
    \boldsymbol{\nu}_7 &= (1,1,1) .
  \end{aligned}
\end{equation}
}\noindent
The first three MIs are the tadpoles with masses $m_1$, $m_2$, $m_3$, respectively; then, we find the three bubbles corresponding to the squared external momenta $p_1^2$, $s$, $p_2^2$, obtained by pinching one denominator at a time; finally, the last MI is the full-scale triangle integral.

We perform the following tests:

\begin{itemize}
\item 
  We start by computing a boundary in a phase-space point $P_1$ with $s = 50$, $p_1^2 = 2$, $p_2^2 = -1/3$, $m_1^2 = 5$, $m_2^2 = 7$, $m_3^2 = 10$ in two independent ways, that is, using both EBR in the limit of vanishing external kinematics $s,\, p_1^2,\, p_2^2 \to 0$ and AMF$^0$, finding perfect agreement on all the required digits.
  With EBR, the only necessary input is the value of the massive tadpoles, whose analytic formula is available in \Line.
  Then, the leading coefficients for the bubbles and the triangle are obtained in terms of the tadpoles by solving the DEs around the regular singular point with vanishing kinematics and imposing the regularity of the solution in such a point.

\item
  Next, we propagate on a line connecting $P_1$ to point $P_2$ with $s = 1$, $p_1^2 = 2$, $p_2^2 = -1/3$, $m_1^2 = 10$, $m_2^2 = 10$, $m_3^2 = 10$, crossing the branch point whose Cutkosky invariant is $z = s-(m_1+m_2)^2$.
  As a consistency check, we also use AMF$^0$ to compute the value of the triangle integral in $P_2$, finding perfect agreement.
\item 
  From $P_1$ we also go to point $P_3$ with $s = 1$, $p_1^2 = 2$, $p_2^2 = -1/3$, and complex masses $m_1^2 = 1 -i$, $m_2^2 = 8/3 -2 i$, $m_3^2 = 17 - i / 4$.
\item
  Finally, we move from $P_3$ to the one-scale regular singular point $P_4$ with $s=-1$, $p_1^2 = 0$, $p_2^2 = 0$, $m_1^2 = 0$, $m_2^2 = 0$, $m_3^2 = 0$.
  The final result is in agreement with the analytic formula,
\begin{equation}
  F^{\boldsymbol{\nu}_7}(\epsilon, s,0,0,0,0,0) = \frac{1}{\epsilon^2}\frac{\Gamma(1+\epsilon)\Gamma^2(1-\epsilon)}{\Gamma(1-2\epsilon)}\frac{(-s)^{-\epsilon}}{s} \,.
\end{equation}
  Furthermore, we observe internal consistency by computing the one-scale triangle with AMF$^0$.
  
\end{itemize}

\noindent
The numerical results for the triangle integral are shown in Table~\ref{tab:1L-triangle-full}.

\begin{table}
\begin{center}
\resizebox{15cm}{!}{
\setlength{\tabcolsep}{5pt}
\begin{tabular}{ccccc}
\toprule 
\toprule 
\tworow{\textbf{target}} &
\tworow{\textbf{$P_1$}} &
\tworow{\textbf{$P_2$}} &
\tworow{\textbf{$P_3$}} &
\tworow{\textbf{$P_4$}} \\ \\ \midrule
\tworow{\textbf{from}} &
\tworow{AMF$^0$, EBR} &
\tworow{AMF$^0$, $P_1$} &
\tworow{$P_1$} &
\tworow{AMF$^0$, $P_3$} \\ \\ \midrule[\heavyrulewidth]
\tworow{$\epsilon^{-2}$} &
\tworow{0} & 
\tworow{0} &
\tworow{0} &
\tworow{\tabnum{-1.000000000000000e0}} \\ \\ \midrule 
\tworow{$\epsilon^{-1}$} &
\tworow{0} & 
\tworow{0} &
\tworow{0} &
\tworow{\tabnum{\!+5.772156649015329e-1}} \\ \\ \midrule 
\tworow{$\epsilon^{0}$} &
\tabnum{-7.599624851460716e-2} & 
\tworow{\tabnum{-5.114624184386078e-2}} &
\tabnum{-9.105983456552547e-2} &
\tworow{\tabnum{\!+6.558780715202539e-1}} \\ 
&
\tabnum{-1.024202715501841e-1*i} & &
\tabnum{-3.405963008295366e-2*i} & \\ \midrule 
\tworow{$\epsilon^{1}$} &
\tabnum{\!+2.851448508579519e-1} &
\tworow{\tabnum{\!+1.461267744725764e-1}} &
\tabnum{\!+2.054866656214297e-1} &
\tworow{\tabnum{\!+2.362111171285093e0}}  \\
&
\tabnum{\!+1.498241156232269e-1*i} & &
\tabnum{\!+2.780936409230585e-2*i} & \\
\midrule 
\tworow{$\epsilon^{2}$} &
\tabnum{-4.359339557414683e-1} &
\tworow{\tabnum{-2.508159227043435e-1}} &
\tabnum{-3.033284294289876e-1} &
\tworow{\tabnum{\!+1.692738940537638e0}} \\
&
\tabnum{-7.119426049903811e-2*i} & &
\tabnum{-2.327298560596528e-2*i} & \\ \midrule 
\tworow{$\epsilon^{3}$} &
\tabnum{\!+4.673966245020759e-1} &
\tworow{\tabnum{\!+3.394894906445344e-1}} &
\tabnum{\!+3.792260921703711e-1} &
\tworow{\tabnum{\!+2.728361494345973e0}} \\
&
\tabnum{\!+5.243128182287680e-3*i} & &
\tabnum{\!+1.589606675868420e-2*i} & \\ \midrule 
\tworow{$\epsilon^{4}$} &
\tabnum{-4.703087868710451e-1} &
\tworow{\tabnum{-4.033919909274164e-1}} &
\tabnum{-4.294046913943785e-1} &
\tworow{\tabnum{\!+1.673348221588670e0}} \\
&
\tabnum{\!+4.807793030293406e-3*i} & &
\tabnum{-9.903139892953955e-3*i} & \\ \bottomrule \bottomrule 
\end{tabular}
}
\end{center}
\caption{Coefficients of the Laurent expansion in $\epsilon$ for the
  one-loop triangle of Eq.\,\eqref{eq:1L-triangle-full-basis} in different phase-space points.
  The heading row indicates the target points whose kinematics is listed in the text.
  The second row specifies for each propagation the starting point or the method used to get boundaries.
  Each entry in subsequent rows corresponds to the
  coefficients associated with the power of $\epsilon$ listed in the
  first column.
  When multiple starting points are indicated, all the
  propagations yield the same numerical results.}
\label{tab:1L-triangle-full}
\end{table}

\subsection{One-loop massless box}
\label{sec:1L-box}
Consider the family of the one-loop massless box:
\begin{equation}
  \label{eq:1L-box}
  F^{\boldsymbol{\nu}} (\epsilon, s, t) = \int_{k} \frac{1}{D_1^{\nu_1} D_2^{\nu_2} D_3^{\nu_3} D_4^{\nu_4}}
\end{equation}
with squared external momenta $s = (p_1+p_2)^2$, $t = (p_2+p_3)^2$, $p_1^2 = p_2^2 = p_3^2 = p_4^2 = 0$ and inverse propagators
\begin{align*}
  D_1 &= k^2 , \\
  D_2 &= (k + p_1)^2 , \\
  D_3 &= (k + p_1 + p_2)^2 , \\
  D_4 &= (k + p_1 + p_2 + p_3)^2 . \numthis
\end{align*}
One can choose a basis $I_i = F^{\boldsymbol{\nu}_i}$ of 3 MIs selecting the $s$- and $t$-channel massless bubbles and the box itself, that is,
{
\allowdisplaybreaks[0]
\small
\begin{align*}
  \label{eq:1L-box-basis}
  \boldsymbol{\nu}_1 &= (1,0,1,0) , \\
  \boldsymbol{\nu}_2 &= (0,1,0,1) , \\
  \boldsymbol{\nu}_3 &= (1,1,1,1) \numthis .
\end{align*}
}

We generate a boundary in point $P_1$ with $s = 1$, $t = -3$ using EBR
in the limit $u = (p_1+p_3)^2 = -s -t \to 0$.  In particular, we solve
the DEs around the singular point $s=1$, $t=-1$, use the analytical
formula for the massless bubbles and get the leading coefficient of
the box by imposing the regularity of the solution in $u=0$.  The
result is in agreement with a boundary generated with AMF$^0$, whose extended basis counts 6 MIs.

We propagate from $P_1$ to point $P_2$ with $s=-11$, $t=5$ crossing
the branch points of Cutkosky invariants $z_1 = s$, $z_2 = t$, finding consistency with AMF$^0$ performed in $P_2$.

Numerical results for the box are shown in Table~\ref{tab:1L-box}.

\begin{table}
\begin{center}
\resizebox{8.18cm}{!}{
\setlength{\tabcolsep}{5pt}
\begin{tabular}{ccc}
\toprule 
\toprule 
\tworow{\textbf{target}} &
\tworow{\textbf{$P_1$}} &
\tworow{\textbf{$P_2$}} \\ \\ \midrule
\tworow{\textbf{from}} &
\tworow{AMF$^0$, EBR} &
\tworow{AMF$^0$, $P_1$} \\ \\ \midrule[\heavyrulewidth]
\tworow{$\epsilon^{-2}$} &
\tworow{\tabnum{-1.333333333333333e0}} &
\tworow{\tabnum{-7.272727272727273e-2}} \\ \\ \midrule 
\tworow{$\epsilon^{-1}$} &
\tabnum{\!+1.502029078980784e0} & 
\tabnum{\!+1.877005278194741e-1} \\ 
&
\tabnum{-2.094395102393195e0*i} &
\tabnum{-1.142397328578107e-1*i} \\ \midrule 
\tworow{$\epsilon^{0}$} &
\tabnum{\!+3.741614747275086e0} & 
\tabnum{\!+2.698156090946971e-3} \\ 
&
\tabnum{\!+3.509845858409871e0*i} &
\tabnum{\!+3.398758787451875e-1*i} \\ \midrule 
\tworow{$\epsilon^{1}$} &
\tabnum{-2.706665331892672e0} & 
\tabnum{-2.846794253590710e-1} \\ 
&
\tabnum{\!+5.235878433110419e0*i} &
\tabnum{-1.143352529230017e-1*i} \\ \midrule 
\tworow{$\epsilon^{2}$} &
\tabnum{-5.048478376080319e0} & 
\tabnum{\!+9.893611975701797e-2} \\ 
&
\tabnum{-1.796965802540394e0*i} &
\tabnum{-1.978243414027738e-1*i} \\ \midrule 
\tworow{$\epsilon^{3}$} &
\tabnum{\!+6.051530711191679e-1} & 
\tabnum{\!+1.402991837463381e-1} \\ 
&
\tabnum{-7.108042701350626e0*i} &
\tabnum{-3.176949541572250e-2*i} \\ \midrule 
\tworow{$\epsilon^{4}$} &
\tabnum{\!+6.960674788336404e0} & 
\tabnum{\!+1.001382259037354e-1} \\ 
&
\tabnum{-6.425634195584692e0*i} &
\tabnum{+7.729488085430293e-3*i} \\ \bottomrule \bottomrule 
\end{tabular}
}
\end{center}
\caption{Coefficients of the Laurent expansion in $\epsilon$ for the
  one-loop massless box of Eq.\,\eqref{eq:1L-box-basis} in different phase-space points.
  The meaning of each entry is the same as for Table~\ref{tab:1L-triangle-full}.  }
\label{tab:1L-box}
\end{table}

\subsection{Full-scale sunrise}
\label{sec:2L-sun-full}
\begin{figure}[ht]
  \centering
  \includegraphics[width=0.8\textwidth]{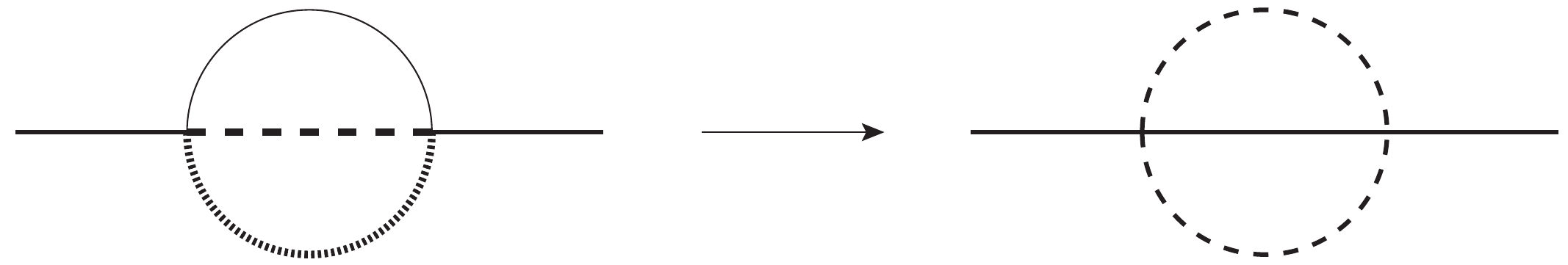}
  \caption{\label{fig:2L-sun-full} (Left) Sunrise diagram with four
    different scales (phase-space point $P_1$ in the text).
    (Right) Mass parameter are varied up to the configuration with two equal internal masses and the third internal mass equal to the external invariant (phase-space point~$P_3$ in the text).}
\end{figure}

Let us consider the family of Feynman integrals associated with the sunrise in Figure~\ref{fig:2L-sun-full},
with incoming squared momentum $s = p^2$ and three different masses $m_1$, $m_2$, $m_3$:
\begin{equation}
  \label{eq:2L-sun-full}
  F^{\boldsymbol{\nu}} (\epsilon, s, m_1^2, m_2^2, m_3^2) = \int_{k_1,k_2} \frac{D_4^{-\nu_4} D_5^{-\nu_5}}{D_1^{\nu_1} D_2^{\nu_2} D_3^{\nu_3} } ,
\end{equation}
with inverse propagators
\begin{equation}
  \allowdisplaybreaks
  \begin{aligned}
    D_1 &= k_1^2 - m_1^2 , &
    D_2 &= (k_2 - p)^2 - m_2^2, \\
    D_3 &= (k_1 + k_2)^2  - m_3^2, &
    D_4 &= k_2^2 , \\
    D_5 &= (k_1 + p)^2 .
  \end{aligned}
\end{equation}
Here, $D_4$ and $D_5$ are auxiliary inverse propagators introduced to complete the basis of scalar products.
The selected basis $I_i = F^{\boldsymbol{\nu}_i}$ consists of 7 MIs with the following powers:
{
\small
\begin{equation}
  \label{eq:2L-sun-full-basis}
  \allowdisplaybreaks
  \begin{aligned}
    \boldsymbol{\nu}_1 &= (1, 1, 0, 0, 0) , &\quad
    \boldsymbol{\nu}_2 &= (1, 0, 1, 0, 0) , \\
    \boldsymbol{\nu}_3 &= (0, 1, 1, 0, 0) , &\quad
    \boldsymbol{\nu}_4 &= (1, 1, 1, -2, 0) , \\
    \boldsymbol{\nu}_5 &= (1, 1, 1, -1, 0) , &\quad
    \boldsymbol{\nu}_6 &= (1, 1, 1, 0, -1) , \\
    \boldsymbol{\nu}_7 &= (1, 1, 1, 0, 0) .
  \end{aligned}
\end{equation}
}

We solve DEs around $s = 0$, $m_1^2 = 10$, $m_2^2 = 10$, $m_3^2 = 10$ and use EBR by imposing the regularity of the solution in the expansion point.
We then propagate to point $P_1$ with $s = -1$, $m_1^2 = 2$, $m_2^2 = 3$, $m_3^2 = 5$, where we find agreement with an AMF$^0$ propagation.
The branch point of $z = s - (m_1 + m_2 + m_3)^2$ is crossed in the propagation from $P_1$ to point $P_2$, 
going from $s=-1$ to $s = 60$ with fixed masses.
From $P_1$ we also propagate to point $P_3$ with $m^2 = -s = 1$, $m_2^2 = m_3^2 = 5$ and from $P_3$ we go to the massless point $P^4$ with $s = -1$.
We also use AMF$^0$ in both $P_2$ and $P_4$, finding agreement on all the required digits.

The numerical results for the sunrise are shown in Table~\ref{tab:2L-sun-full}.

\begin{table}
\begin{center}
\resizebox{15cm}{!}{
\setlength{\tabcolsep}{5pt}
\begin{tabular}{ccccc}
\toprule \toprule 
\tworow{\textbf{target}} &
\tworow{\textbf{$P_1$}} &
\tworow{\textbf{$P_2$}} &
\tworow{\textbf{$P_3$}} &
\tworow{\textbf{$P_4$}} \\ \\ \midrule
\tworow{\textbf{from}} &
\tworow{AMF$^0$, EBR} &
\tworow{AMF$^0$, $P_1$} &
\tworow{$P_1$} &
\tworow{AMF$^0$, $P_3$} \\ \\ \midrule[\heavyrulewidth]
\tworow{$\epsilon^{-4}$} &
\tworow{0} & 
\tworow{0} &
\tworow{0} &
\tworow{0} \\ \\ \midrule 
\tworow{$\epsilon^{-3}$} &
\tworow{0} & 
\tworow{0} &
\tworow{0} &
\tworow{0} \\ \\ \midrule 
\tworow{$\epsilon^{-2}$} &
\tworow{\tabnum{\!+5.000000000000000e0}} &
\tworow{\tabnum{\!+5.000000000000000e0}} &
\tworow{\tabnum{\!+5.500000000000000e0}} &
\tworow{0} \\ \\ \midrule 
\tworow{$\epsilon^{-1}$} &
\tworow{\tabnum{-3.251477438310050e0}} &
\tworow{\tabnum{-1.850147743831005e1}} &
\tworow{\tabnum{-5.693751438257865e0}} &
\tworow{\tabnum{\!+2.500000000000000e-1}} \\ \\ \midrule 
\tworow{$\epsilon^{0}$} &
\tworow{\tabnum{\!+1.188378767646979e1}} &
\tabnum{\!+6.552872234370230e1} &
\tworow{\tabnum{\!+1.867337540448070e1}} &
\tworow{\tabnum{\!+1.336392167549234e0}} \\
& &
\tabnum{-1.758371010882413e1*i} & & \\ \midrule 
\tworow{$\epsilon^{1}$} &
\tworow{\tabnum{\!+1.952137703514755e1}} &
\tabnum{-1.091156083475895e2} &
\tworow{\tabnum{\!+4.626131138234519e0}} &
\tworow{\tabnum{\!+5.066904534261821e0}} \\
& &
\tabnum{\!+2.967183417356042e1*i} & & \\ \midrule 
\tworow{$\epsilon^{2}$} &
\tworow{\tabnum{-2.160341605441262e1}} &
\tabnum{\!+3.251877471184126e2} &
\tworow{\tabnum{\!+7.465797730320954e0}} &
\tworow{\tabnum{\!+1.434873581897869e1}} \\
& &
\tabnum{-1.125285386030628e1*i} & & \\ \bottomrule \bottomrule 
\end{tabular}
}
\end{center}
\caption{Coefficients of the Laurent expansion in $\epsilon$ for the sunrise integral of Eq.\,\eqref{eq:2L-sun-full-basis} at several phase-space points.
  The meaning of each entry is the same as for Table~\ref{tab:1L-triangle-full}.
}
\label{tab:2L-sun-full}
\end{table}

\subsection{Two-loop non-planar triangle with mass}
\label{sec:2L-QEDFFb}
Let $F^{\boldsymbol{\nu}}(\epsilon, s, m^2)$ be the family of the two-loop three-point function in Figure~\ref{fig:2L-QEDFFb}:
\begin{figure}[ht]
  \centering
  \includegraphics[width=0.7\textwidth]{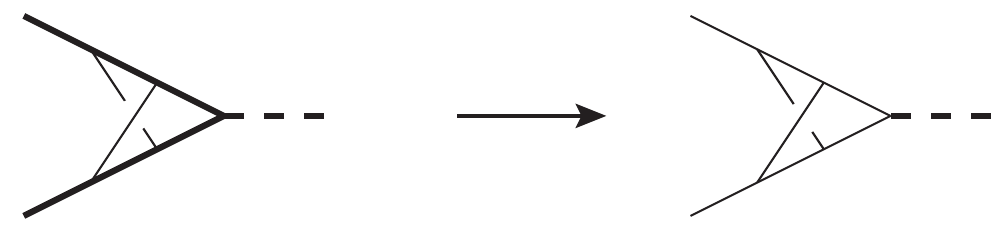}
  \caption{\label{fig:2L-QEDFFb} (Left) Non-planar triangle diagram with a common internal and external mass highlighted with a bold line (phase-space point $P_1$ in the text).
    (Right) The same diagram in the massless limit (phase-space point $P_3$ in the text).}
\end{figure}
\begin{equation}
  \label{eq:2L-QEDFFb}
  F^{\boldsymbol{\nu}} (\epsilon, s, m^2) = \int_{k_1,k_2} \frac{D_7^{-\nu_7}}{D_1^{\nu_1} D_2^{\nu_2} D_3^{\nu_3} D_4^{\nu_4} D_5^{\nu_5} D_6^{\nu_6}}
\end{equation}
with squared external momenta $s = (p_1+p_2)^2$, $m^2= p_1^2 = p_2^2$ and inverse propagators
\begin{equation}
  \allowdisplaybreaks
  \begin{aligned}
    D_1 &= k_1^2 , &
    D_2 &= k_2^2 , \\
    D_3 &= (k_1 - p_1)^2 - m^2 , &
    D_4 &= (k_2 + p_2)^2 - m^2 , \\
    D_5 &= (k_1 + k_2 - p_1)^2 - m^2 , &
    D_6 &= (k_1 + k_2 + p_2)^2 - m^2 , \\
    D_7 &= (k_1 + k_2)^2 .
  \end{aligned}
\end{equation}
Here, $D_7$ is a numerator that appears only in one out of the 16 MIs of the selected basis:
{
\small
\begin{equation}
  \label{eq:2L-QEDFFb-basis}
  \allowdisplaybreaks
  \begin{aligned}
    \boldsymbol{\nu}_1 &= (0, 0, 1, 1, 0, 0, 0) , &
    \boldsymbol{\nu}_2 &= (1, 1, 0, 0, 1, 0, 0) , \\
    \boldsymbol{\nu}_3 &= (1, -1, 0, 1, 1, 0, 0) , &
    \boldsymbol{\nu}_4 &= (1, 0, 0, 1, 1, 0, 0) , \\
    \boldsymbol{\nu}_5 &= (0, 0, 1, 1, 1, 0, 0) , &
    \boldsymbol{\nu}_6 &= (0, 0, 1, 0, 1, 1, 0) , \\
    \boldsymbol{\nu}_7 &= (1, 1, 0, 1, 1, 0, 0) , &
    \boldsymbol{\nu}_8 &= (1, -1, 1, 1, 1, 0, 0) , \\
    \boldsymbol{\nu}_9 &= (1, 0, 1, 1, 1, 0, 0) , &
    \boldsymbol{\nu}_{10} &= (1, 1, 0, 0, 1, 1, 0) , \\
    \boldsymbol{\nu}_{11} &= (-1, 0, 1, 1, 1, 1, 0) , &
    \boldsymbol{\nu}_{12} &= (0, 0, 1, 1, 1, 1, 0) , \\
    \boldsymbol{\nu}_{13} &= (1, 1, 1, -1, 1, 1, 0) , &
    \boldsymbol{\nu}_{14} &= (1, 1, 1, 0, 1, 1, 0) , \\
    \boldsymbol{\nu}_{15} &= (1, 1, 1, 1, 1, 1, -1) , &
    \boldsymbol{\nu}_{16} &= (1, 1, 1, 1, 1, 1, 0) .
  \end{aligned}
\end{equation}
}

We use AMF$^0$ with an extended basis of 52 MIs to generate a boundary for point $P_1$ with $s = 10$, $m^2 = 1$ and propagate to point $P_2$ with $s = 1$, $m^2 = 3$, crossing the branch point with Cutkosky invariant $z = s - 4 m^2$.
We then go from $P_2$ to point $P_3$ in the massless limit.
We use AMF$^0$ also in $P_3$, finding agreement up to the required number of digits.
The numerical results for the last MI $F^{\boldsymbol{\nu}_{16}}$ are shown in Table~\ref{tab:2L-QEDFFb}.

\begin{table}
\begin{center}
\resizebox{11.59cm}{!}{
\setlength{\tabcolsep}{5pt}
\begin{tabular}{cccc}
\toprule \toprule 
\tworow{\textbf{target}} &
\tworow{\textbf{$P_1$}} &
\tworow{\textbf{$P_2$}} &
\tworow{\textbf{$P_3$}} \\ \\ \midrule
\tworow{\textbf{from}} &
\tworow{AMF$^0$} &
\tworow{$P_1$} &
\tworow{AMF$^0$, $P_2$} \\ \\ \midrule[\heavyrulewidth]
\tworow{$\epsilon^{-4}$} &
\tworow{0} & 
\tworow{0} &
\tworow{\tabnum{\!+1.000000000000000e0}} \\ \\ \midrule 
\tworow{$\epsilon^{-3}$} &
\tworow{0} & 
\tworow{0} &
\tabnum{-1.154431329803066e0} \\
& & &
\tabnum{\!+6.283185307179586e0*i} \\ \midrule 
\tworow{$\epsilon^{-2}$} &
\tworow{0} & 
\tworow{0} &
\tabnum{-2.894245735565264e1} \\
& & &
\tabnum{-7.253505969566414e0*i} \\ \midrule 
\tworow{$\epsilon^{-1}$} &
\tabnum{\!+2.532501153536048e-1} &
\tworow{\tabnum{-3.058450755305179e-2}} &
\tabnum{\!+6.680132569623135e-1} \\
&
\tabnum{\!+1.376560680870821e-1*i} &
&
\tabnum{-9.916741832990889e1*i}  \\ \midrule 

\tworow{$\epsilon^{0}$} &
\tabnum{-1.137868788629137e0} &
\tworow{\tabnum{\!+6.882432933483959e-2}} &
\tabnum{\!+2.306015883275194e2} \\
&
\tabnum{\!+1.315450793632957e0*i} &
&
\tabnum{-9.125506150626736e1*i}  \\ \midrule 

\tworow{$\epsilon^{1}$} &
\tabnum{-5.535444498587951e0} &
\tworow{\tabnum{\!+5.232509250247894e-2}} &
\tabnum{\!+4.317677285401460e2} \\
&
\tabnum{-1.578608277056101e0*i} &
&
\tabnum{\!+3.615355918032282e2*i} \\ \midrule 
\tworow{$\epsilon^{2}$} &
\tabnum{-1.199497745643981e1 } &
\tworow{\tabnum{\!+8.195254040212031e-1}} &
\tabnum{\!+1.850496772277360e1} \\
&
\tabnum{-8.780073080609521e0*i} &
&
\tabnum{\!+1.260787755350661e3*i} \\ \bottomrule \bottomrule 
\end{tabular}
}
\end{center}
\caption{Coefficients of the Laurent expansion in $\epsilon$ for the two-loop non-planar triangle integral $F^{\boldsymbol{\nu}_{16}}$ of Eq.\,\eqref{eq:2L-QEDFFb-basis} in different phase-space points.
  The meaning of each entry is the same as for Table~\ref{tab:1L-triangle-full}.
}
\label{tab:2L-QEDFFb}
\end{table}

\subsection{Two-loop planar box with a massive loop}
\label{sec:2L-box-ml-m}
Let us consider the integral family of the two-loop planar box of Figure~\ref{fig:2L-box-ml-m}:
\begin{figure}[ht]
  \centering
  \includegraphics[width=1\textwidth]{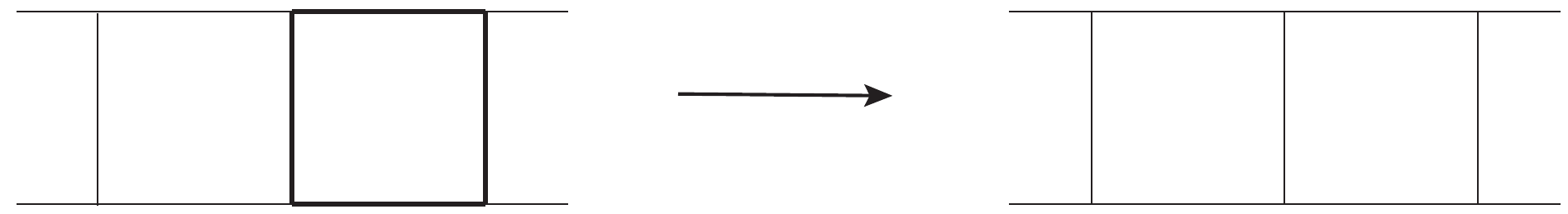}
  \caption{\label{fig:2L-box-ml-m} (Left) Double box diagram with a massive loop highlighted with a bold line (phase-space point $P_1$ in the text).
    (Right) The same diagram in the massless limit (phase-space point $P_3$ in the text).}
\end{figure}
\begin{equation}
  \label{eq:2L-box-ml-m}
  F^{\boldsymbol{\nu}} (\epsilon, s, t, m^2) = \int_{k_1,k_2} \frac{D_8^{-\nu_8} D_9^{-\nu_9}}{D_1^{\nu_1} D_2^{\nu_2} D_3^{\nu_3} D_4^{\nu_4} D_5^{\nu_5} D_6^{\nu_6} D_7^{\nu_7}}
\end{equation}
with squared external momenta $s = (p_1+p_2)^2$, $t = (p_2+p_3)^2$, $p_1^2 = p_2^2 = p_3^2 = p_4^2 = 0$ and inverse propagators
\begin{equation}
  \allowdisplaybreaks
  \begin{aligned}
    D_1 &= k_1^2 , &
    D_2 &= (k_1 - p_1)^2 , \\
    D_3 &= (k_1 + p_2)^2 , &
    D_4 &= (k_2 - p_2)^2 - m^2 , \\
    D_5 &= (k_2 + p_1)^2 - m^2 , &
    D_6 &= (k_1 + k_2)^2 - m^2 , \\
    D_7 &= (k_2 - p_2 - p_3)^2 - m^2 , &
    D_8 &= k_2^2 , \\
    D_9 &= (k_1 + p_3)^2 ,
  \end{aligned}
\end{equation}
where $D_8$ and $D_9$ are used to complete the basis of scalar products.
The selected basis $I_i = F^{\boldsymbol{\nu}_i}_i$ has 32 MIs:
{
\small
\allowdisplaybreaks
  \begin{align*}
    \label{eq:2L-box-ml-m-basis}
    \boldsymbol{\nu}_1 &= (0, 0, 0, 1, 0, 1, 0, 0, 0) , &\!\!
    \boldsymbol{\nu}_2 &= (0, 1, 1, 1, 0, 0, 0, 0, 0) , &\!\!
    \boldsymbol{\nu}_3 &= (-1, 1, 0, 1, 0, 1, 0, 0, 0) , \\
    \boldsymbol{\nu}_4 &= (0, 1, 0, 1, 0, 1, 0, 0, 0) , &\!\!
    \boldsymbol{\nu}_5 &= (0, 0, 0, 1, 1, 1, 0, 0, 0) , &\!\!
    \boldsymbol{\nu}_6 &= (1, -1, 0, 0, 0, 1, 1, 0, 0) , \\
    \boldsymbol{\nu}_7 &= (1, 0, 0, 0, 0, 1, 1, 0, 0) , &\!\!
    \boldsymbol{\nu}_8 &= (0, 1, 1, 1, 1, 0, 0, 0, 0) , &\!\!
    \boldsymbol{\nu}_9 &= (1, -1, 0, 1, 1, 1, 0, 0, 0) , \\
    \boldsymbol{\nu}_{10} &= (1, 0, 0, 1, 1, 1, -1, 0, 0) , &\!\!
    \boldsymbol{\nu}_{11} &= (1, 0, 0, 1, 1, 1, 0, 0, 0) , &\!\!
    \boldsymbol{\nu}_{12} &= (-1, 1, 1, 0, 0, 1, 1, 0, 0), \\
    \boldsymbol{\nu}_{13} &= (0, 1, 1, 0, 0, 1, 1, 0, 0) , &\!\!
    \boldsymbol{\nu}_{14} &= (1, 0, 0, 1, 0, 1, 1, 0, 0) , &\!\!
    \boldsymbol{\nu}_{15} &= (0, 1, 0, 1, 0, 1, 1, 0, 0) , \\
    \boldsymbol{\nu}_{16} &= (0, 0, 0, 1, 1, 1, 1, 0, 0) , &\!\!
    \boldsymbol{\nu}_{17} &= (0, 1, 1, 1, 1, 1, 0, 0, 0) , &\!\!
    \boldsymbol{\nu}_{18} &= (0, 1, 1, 1, 1, 0, 1, 0, 0) , \\
    \boldsymbol{\nu}_{19} &= (1, 1, 1, -1, 0, 1, 1, 0, 0) , &\!\!
    \boldsymbol{\nu}_{20} &= (1, 1, 1, 0, 0, 1, 1, 0, 0) , &\!\!
    \boldsymbol{\nu}_{21} &= (1, 1, -1, 1, 0, 1, 1, 0, 0) , \\
    \boldsymbol{\nu}_{22} &= (1, 1, 0, 1, 0, 1, 1, 0, 0) , &\!\!
    \boldsymbol{\nu}_{23} &= (0, 1, 1, 1, 0, 1, 1, 0, 0) , &\!\!
    \boldsymbol{\nu}_{24} &= (1, -1, 0, 1, 1, 1, 1, 0, 0) , \\
    \boldsymbol{\nu}_{25} &= (1, 0, 0, 1, 1, 1, 1, -1, 0) , &\!\!
    \boldsymbol{\nu}_{26} &= (1, 0, 0, 1, 1, 1, 1, 0, 0) , &\!\!
    \boldsymbol{\nu}_{27} &= (1, 1, 1, 1, 0, 1, 1, 0, 0) , \\
    \boldsymbol{\nu}_{28} &= (0, 1, 1, 1, 1, 1, 1, 0, 0) , &\!\!
    \boldsymbol{\nu}_{29} &= (1, 1, 1, 1, 1, 1, 1, -2, 0) , &\!\!
    \boldsymbol{\nu}_{30} &= (1, 1, 1, 1, 1, 1, 1, -1, 0) , \\
    \boldsymbol{\nu}_{31} &= (1, 1, 1, 1, 1, 1, 1, 0, -1) , &\!\!
    \boldsymbol{\nu}_{32} &= (1, 1, 1, 1, 1, 1, 1, 0, 0) \numthis .
\end{align*}
}

We start by looking for a boundary in the limit of vanishing external kinematics $s, t \to 0$ at $m^2 = 10$.
Around this regular singular point, EBR can be used to analyze the regions where the two loop momenta $k_1$ and $k_2$ are negligible or not w.r.t. $m^2$.
The first MI behaves like a squared tadpole and the second one like a product of a tadpole and a massless bubble.
Their leading coefficients are thus obtained with the analytic formulae implemented in \Line.
For the other MIs, when scale-less contributions are set to zero, only two kinds of regions survive: they have power behavior $\eta^{\lambda}$ with $\lambda = 0$ or $\lambda=n - \epsilon$, where $n \in \mathbb{Z}$ is an integer that can vary from one master to another.

For instance, the last MI behaves like
\begin{equation}
  F^{\boldsymbol{\nu}_{32}}(\epsilon, s, m^2) \widesim[2]{s, t \to 0}
  \includegraphics[width=0.44\textwidth,valign=c]{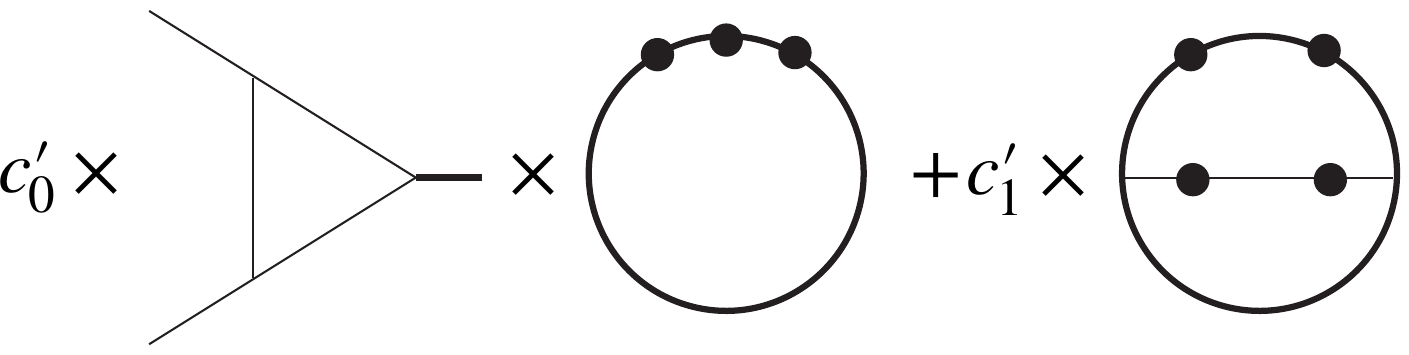}
  = c_0 + c_1 \eta^{-1-\epsilon} .
\end{equation}
The offset $n = -1$ can be obtained by counting the energy dimension of the massless triangle. In fact, since the latter only has the scale $\eta$, with one loop and three inverse propagator powers, it must be proportional to $\eta^{\lambda}$ with $\lambda = d/2 - 3 = - 1 - \epsilon$.
For this example, there is no need to manually compute the actual value of the coefficients $c_0$ and $c_1$: we only need to know what the power behaviors are.
In fact, imposing that the series solution around the regular singular point has no other power behavior besides $\lambda = 0$ and $\lambda= -1 - \epsilon$, we can obtain the coefficients for the solutions of the associated homogeneous equation and build the complete solution that can be evaluated in $s, t \ne 0$.
For instance, we choose point $P_1$ with $s = -1$, $t = 2$, finding perfect agreement with an AMF$^{0}$ run involving an extended basis of 68 MIs.

Having completed the outlined procedure, we cross the branch points of $z_1 =  s$, $z_2 = s - 4 m^2$ and $z_3 = t - 4 m^2$, moving from $P_1$ to point $P_2$ with $s = 70$, $t = 50$, $m^2 = 10$.
Then, we push the mass down to zero in $P_3$, in full agreement with an AMF$^0$ propagation directly performed on the massless box with $s = 70$, $t = 50$.
The numerical results for the two-loop box $F^{\boldsymbol{\nu}_{32}}$ are shown in Table~\ref{tab:2L-box-ml-m}.

\begin{table}
\begin{center}
\resizebox{11.59cm}{!}{
\setlength{\tabcolsep}{5pt}
\begin{tabular}{cccc}
\toprule \toprule 
\tworow{\textbf{target}} &
\tworow{\textbf{$P_1$}} &
\tworow{\textbf{$P_2$}} &
\tworow{\textbf{$P_3$}} \\ \\ \midrule
\tworow{\textbf{from}} &
\tworow{AMF$^0$, EBR} &
\tworow{$P_1$} &
\tworow{AMF$^0$, $P_2$} \\ \\ \midrule[\heavyrulewidth]
\tworow{$\epsilon^{-4}$} &
\tworow{0} & 
\tworow{0} &
\tworow{\tabnum{\!+1.632653061224490e-5}} \\ \\ \midrule 
\tworow{$\epsilon^{-3}$} &
\tworow{0} & 
\tworow{0} &
\tabnum{-1.507074533571472e-4} \\
& & &
\tabnum{\!+1.025826172600749e-4*i} \\ \midrule 
\tworow{$\epsilon^{-2}$} &
\tworow{\tabnum{-1.684311982263061e-3}} & 
\tabnum{\!+7.121750612221514e-5} &
\tabnum{\!+2.720746512604996e-4} \\
& &
\tabnum{\!+1.223851404355579e-4*i} &
\tabnum{-9.469228566160803e-4*i} \\ \midrule 
\tworow{$\epsilon^{-1}$} &
\tworow{\tabnum{\!+4.026956116103587e-3}} & 
\tabnum{-7.645333935948279e-4} &
\tabnum{\!+1.572347464421193e-3} \\
& &
\tabnum{-3.758110807119310e-4*i} &
\tabnum{\!+3.059428585636381e-3*i} \\ \midrule 
\tworow{$\epsilon^{0}$} &
\tworow{\tabnum{-3.997722931454625e-3}} & 
\tabnum{\!+1.621191987913520e-3} &
\tabnum{-8.340803170789194e-3} \\
& &
\tabnum{-1.376157443003446e-4*i} &
\tabnum{-2.581654837967916e-3*i} \\ \midrule 
\tworow{$\epsilon^{1}$} &
\tworow{\tabnum{\!+6.237012138664067e-3}} & 
\tabnum{-2.779941041112323e-3} &
\tabnum{\!+1.483674698459523e-2} \\
& &
\tabnum{-3.108819053117712e-5*i} &
\tabnum{-8.593463886823766e-3*i} \\ \midrule 
\tworow{$\epsilon^{2}$} &
\tworow{\tabnum{-4.987777863769356e-3}} & 
\tabnum{\!+5.841649978319638e-3} &
\tabnum{-4.995133665555594e-3} \\
& &
\tabnum{-1.900890782973601e-3*i} &
\tabnum{\!+2.645276326148751e-2*i} \\ \bottomrule \bottomrule 
\end{tabular}
}
\end{center}
\caption{Coefficients of the Laurent expansion in $\epsilon$ for the two-loop planar box $F^{\boldsymbol{\nu}_{32}}$ of Eq.\,\eqref{eq:2L-box-ml-m-basis} in different phase-space points.
  The meaning of each entry is the same as for Table~\ref{tab:1L-triangle-full}.
}
\label{tab:2L-box-ml-m}
\end{table}

\subsection{Two-loop non-planar boxes with mass}
\begin{figure}[ht]
  \includegraphics[width=0.45\textwidth]{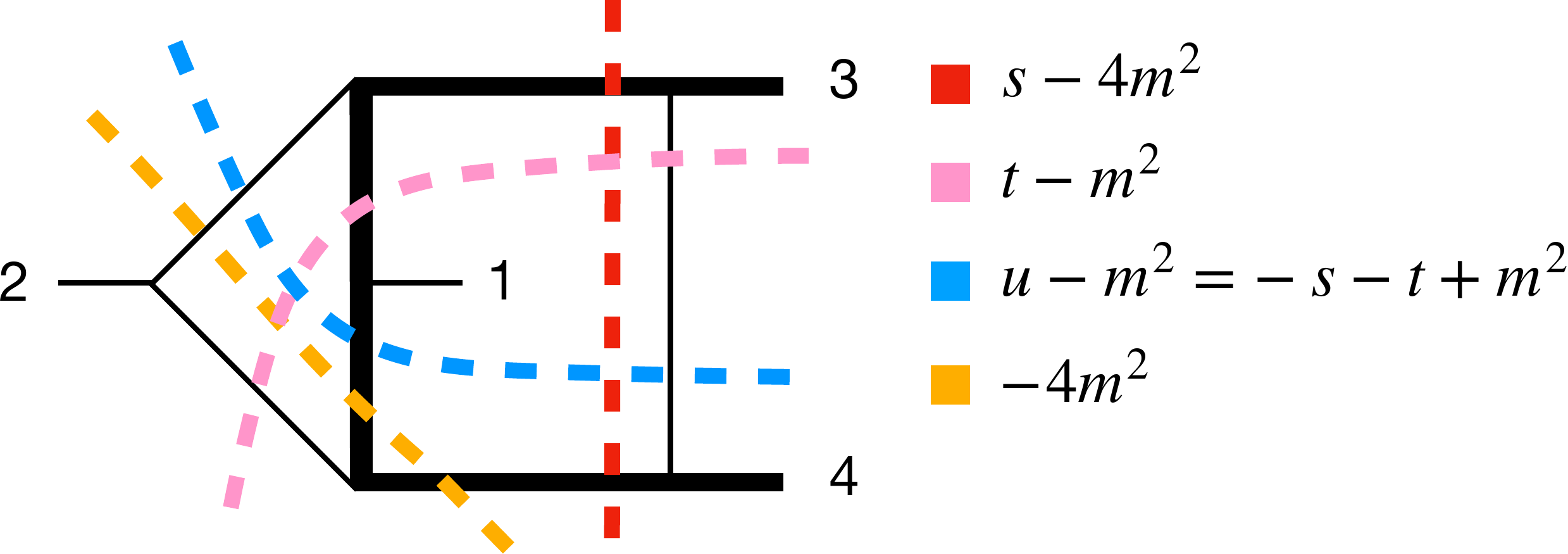}
  \hfill
  \includegraphics[width=0.49\textwidth]{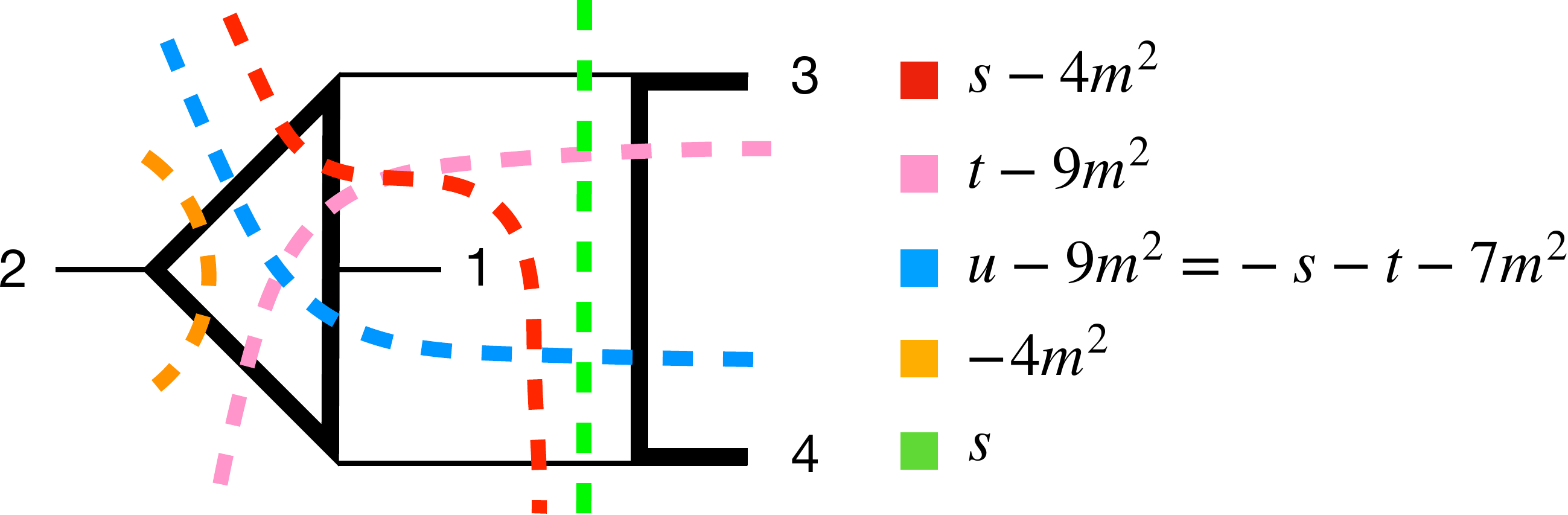}
  \caption{\label{fig:2L-box-np-1m-cuts} Non-planar boxes with one internal and external mass parameter. Cutkosky cuts are displayed with dashed lines and the corresponding invariants are listed in the legends next to each diagram.}
\end{figure}

We consider the integral families of the two non-planar boxes in Figure~\ref{fig:2L-box-np-1m-cuts},
both of which have squared external momenta $s = (p_1+p_2)^2$, $t = (p_2+p_3)^2$, $p_1^2 = p_2^2 = 0$, $p_3^2 = p_4^2 = m^2$:
\begin{equation}
  F^{\boldsymbol{\nu}} (\epsilon, s, t, m^2) = \int_{k_1,k_2} \frac{D_8^{-\nu_8} D_9^{-\nu_9}}{D_1^{\nu_1} D_2^{\nu_2} D_3^{\nu_3} D_4^{\nu_4} D_5^{\nu_5} D_6^{\nu_6} D_7^{\nu_7}}
\end{equation}
with inverse propagators separately specified below for the two cases.

\paragraph{1st box.}
The box in the left panel of Figure~\ref{fig:2L-box-np-1m-cuts} has inverse propagators
\begin{equation}
  \allowdisplaybreaks
  \begin{aligned}
    D_1 &= k_2^2, &\quad
    D_2 &= (k_2 - p_2)^2 , \\
    D_3 &= (k_2 - k_1 - p_2 - p_3)^2 , &\quad
    D_4 &= k_1^2 - m^2 , \\
    D_5 &= (k_1 - p_1)^2 - m^2 , &\quad
    D_6 &= (k_2 - k_1 - p_2)^2 - m^2 , \\
    D_7 &= (k_2 - k_1 + p_1)^2 - m^2 , &\quad
    D_8 &= (k_1 + p_3)^2 , \\
    D_9 &= (k_1 + p_2)^2 .
  \end{aligned}
\end{equation}
We found a basis $I_i = F^{\boldsymbol{\nu}_i}$ with the following 55 MIs:

{
\small
\allowdisplaybreaks
  \begin{align*}
    \label{eq:2L-box-np-1m-basis}
    \boldsymbol{\nu}_1 &= (0, 0, 0, 1, 0, 1, 0, 0, 0) , &\quad
    \boldsymbol{\nu}_2 &= (1, -1, 1, 1, 0, 0, 0, 0, 0) , &\quad
    \boldsymbol{\nu}_3 &= (1, 0, 1, 1, 0, 0, 0, 0, 0) , \\
    \boldsymbol{\nu}_4 &= (0, 1, 1, 1, 0, 0, 0, 0, 0) , &\quad
    \boldsymbol{\nu}_5 &= (-1, 1, 1, 0, 1, 0, 0, 0, 0) , &\quad
    \boldsymbol{\nu}_6 &= (0, 1, 1, 0, 1, 0, 0, 0, 0) , \\
    \boldsymbol{\nu}_7 &= (1, -1, 0, 0, 1, 1, 0, 0, 0) , &\quad
    \boldsymbol{\nu}_8 &= (1, 0, 0, 0, 1, 1, 0, 0, 0) , &\quad
    \boldsymbol{\nu}_9 &= (0, 0, 0, 1, 0, 1, 1, 0, 0) , \\
    \boldsymbol{\nu}_{10} &= (1, -1, 1, 1, 0, 1, 0, 0, 0) , &\quad
    \boldsymbol{\nu}_{11} &= (1, 0, 1, 1, 0, 1, 0, 0, 0) , &\quad
    \boldsymbol{\nu}_{12} &= (1, 0, 1, 0, 1, 1, 0, 0, 0) , \\
    \boldsymbol{\nu}_{13} &= (-1, 1, 1, 0, 1, 1, 0, 0, 0) , &\quad
    \boldsymbol{\nu}_{14} &= (0, 1, 1, 0, 1, 1, 0, 0, 0) , &\quad
    \boldsymbol{\nu}_{15} &= (1, 0, 0, 1, 1, 1, 0, 0, 0) , \\
    \boldsymbol{\nu}_{16} &= (1, -1, 0, 1, 0, 1, 1, 0, 0) , &\quad
    \boldsymbol{\nu}_{17} &= (1, 0, -1, 1, 0, 1, 1, 0, 0) , &\quad
    \boldsymbol{\nu}_{18} &= (1, 0, 0, 1, 0, 1, 1, 0, 0) , \\
    \boldsymbol{\nu}_{19} &= (1, 1, 1, 1, 1, -1, 0, 0, 0) , &\quad
    \boldsymbol{\nu}_{20} &= (1, 1, 1, 1, 1, 0, 0, 0, 0) , &\quad
    \boldsymbol{\nu}_{21} &= (1, 1, 1, -1, 1, 1, 0, 0, 0) , \\
    \boldsymbol{\nu}_{22} &= (1, 1, 1, 0, 1, 1, -1, 0, 0) , &\quad
    \boldsymbol{\nu}_{23} &= (1, 1, 1, 0, 1, 1, 0, 0, 0) , &\quad
    \boldsymbol{\nu}_{24} &= (1, -1, 1, 1, 1, 1, 0, 0, 0) , \\
    \boldsymbol{\nu}_{25} &= (1, 0, 1, 1, 1, 1, -1, 0, 0) , &\quad
    \boldsymbol{\nu}_{26} &= (1, 0, 1, 1, 1, 1, 0, -1, 0) , &\quad
    \boldsymbol{\nu}_{27} &= (1, 0, 1, 1, 1, 1, 0, 0, 0) , \\
    \boldsymbol{\nu}_{28} &= (0, 1, 1, 1, 1, 1, 0, 0, 0) , &\quad
    \boldsymbol{\nu}_{29} &= (1, 1, 1, 1, -1, 0, 1, 0, 0) , &\quad
    \boldsymbol{\nu}_{30} &= (1, 1, 1, 1, 0, -1, 1, 0, 0) , \\
    \boldsymbol{\nu}_{31} &= (1, 1, 1, 1, 0, 0, 1, 0, 0) , &\quad
    \boldsymbol{\nu}_{32} &= (1, 0, 1, 1, 1, 0, 1, 0, 0) , &\quad
    \boldsymbol{\nu}_{33} &= (-1, 1, 1, 1, 1, 0, 1, 0, 0) , \\
    \boldsymbol{\nu}_{34} &= (0, 1, 1, 1, 1, -1, 1, 0, 0) , &\quad
    \boldsymbol{\nu}_{35} &= (0, 1, 1, 1, 1, 0, 1, -1, 0) , &\quad
    \boldsymbol{\nu}_{36} &= (0, 1, 1, 1, 1, 0, 1, 0, 0) , \\
    \boldsymbol{\nu}_{37} &= (1, -1, 1, 1, 0, 1, 1, 0, 0) , &\quad
    \boldsymbol{\nu}_{38} &= (1, 0, 1, 1, 0, 1, 1, 0, 0) , &\quad
    \boldsymbol{\nu}_{39} &= (-1, 1, 1, 0, 1, 1, 1, 0, 0) , \\
    \boldsymbol{\nu}_{40} &= (0, 1, 1, 0, 1, 1, 1, 0, 0) , &\quad
    \boldsymbol{\nu}_{41} &= (1, -1, 0, 1, 1, 1, 1, 0, 0) , &\quad
    \boldsymbol{\nu}_{42} &= (1, 0, 0, 1, 1, 1, 1, 0, 0) , \\
    \boldsymbol{\nu}_{43} &= (1, 1, 1, 1, 1, 1, -1, 0, 0) , &\quad
    \boldsymbol{\nu}_{44} &= (1, 1, 1, 1, 1, 1, 0, -1, 0) , &\quad
    \boldsymbol{\nu}_{45} &= (1, 1, 1, 1, 1, 1, 0, 0, 0) , \\
    \boldsymbol{\nu}_{46} &= (1, 1, 1, 1, 1, -1, 1, 0, 0) , &\quad
    \boldsymbol{\nu}_{47} &= (1, 1, 1, 1, 1, 0, 1, -1, 0) , &\quad
    \boldsymbol{\nu}_{48} &= (1, 1, 1, 1, 1, 0, 1, 0, 0) , \\
    \boldsymbol{\nu}_{49} &= (1, 1, 0, 1, 1, 1, 1, 0, 0) , &\quad
    \boldsymbol{\nu}_{50} &= (1, 0, 1, 1, 1, 1, 1, 0, 0) , &\quad
    \boldsymbol{\nu}_{51} &= (0, 1, 1, 1, 1, 1, 1, 0, 0) , \\
    \boldsymbol{\nu}_{52} &= (1, 1, 1, 1, 1, 1, 1, -2, 0) , &\quad
    \boldsymbol{\nu}_{53} &= (1, 1, 1, 1, 1, 1, 1, -1, 0) , &\quad
    \boldsymbol{\nu}_{54} &= (1, 1, 1, 1, 1, 1, 1, 0, -1) , \\
    \boldsymbol{\nu}_{55} &= (1, 1, 1, 1, 1, 1, 1, 0, 0) .
    \numthis
  \end{align*}
}
We generate a boundary in point $Q_1$ with $s = 1$, $t = 2$, $m^2 = 100$ using AMF$^0$ with an extended basis of 144 MIs.
We then propagate from $Q_1$ to point $Q_2$ with $s = 500$, $t = 150$, $m^2 = 100$, crossing the branch points of $z_1 = s - 4 m^2$, $z_2 = t - m^2$, $z_3 = -s - t + m^{2}$, and finding perfect agreement with AMF$^0$ performed in $Q_2$.

The numerical results for the two-loop non-planar box $F^{\boldsymbol{\nu}_{55}}$ are shown in Table~\ref{tab:2L-box-np-1m}.

%
\begin{table}
\begin{center}
\resizebox{15cm}{!}{
\setlength{\tabcolsep}{5pt}
\begin{tabular}{ccccc}
\toprule \toprule 
\tworow{\textbf{target}} &
\tworow{\textbf{$Q_1$}} &
\tworow{\textbf{$Q_2$}} &
\tworow{\textbf{$P_1$}} &
\tworow{\textbf{$P_2$}} \\ \\ \midrule
\tworow{\textbf{from}} &
\tworow{AMF$^0$} &
\tworow{AMF$^0$, $Q_1$} &
\tworow{AMF$^0$} &
\tworow{AMF$^0$, $P_1$} \\ \\ \midrule[\heavyrulewidth]
\tworow{$\epsilon^{-4}$} &
\tworow{0} & 
\tworow{0} &
\tworow{0} &
\tworow{0} \\ \\ \midrule 
\tworow{$\epsilon^{-3}$} &
\tworow{\tabnum{-2.634309928357791e-7}} &
\tabnum{\!+7.825617108436437e-8} &
\tworow{0} &
\tworow{0} \\
& &
\tabnum{-2.554478084014810e-7*i} & & \\ \midrule 
\tworow{$\epsilon^{-2}$} &
\tabnum{\!+2.177434402618331e-6} &
\tabnum{\!+5.136099594647812e-9} &
\tworow{0} &
\tworow{0} \\
&
\tabnum{-1.655185743641498e-6*i} &
\tabnum{\!+3.245051324395477e-6*i} & & \\ \midrule 
\tworow{$\epsilon^{-1}$} &
\tabnum{\!+2.177434402618331e-6} &
\tabnum{\!+5.136099594647812e-9} &
\tworow{0} &
\tworow{0} \\
&
\tabnum{\!+1.533076938553119e-5*i} &
\tabnum{-3.407024087192466e-5*i} & & \\ \midrule 
\tworow{$\epsilon^{0}$} &
\tabnum{-2.810879169233962e-5} &
\tabnum{\!+2.470711494037188e-4} &
\tabnum{\!+2.576938753803745e-1} &
\tabnum{\!+2.518740723653660e-1} \\
&
\tabnum{-3.761642841819541e-5*i} &
\tabnum{-6.343358651146831e-5*i} &
\tabnum{-2.465521721983634e-1*i} &
\tabnum{-1.169079848124980e-1*i} \\ \midrule
\tworow{$\epsilon^{1}$} &
\tabnum{\!+6.424181660342731e-5} & 
\tabnum{\!+3.561272520516187e-5} & 
\tabnum{\!+9.839059948409147e-1} & 
\tabnum{\!+8.377932210850515e-1} \\
&
\tabnum{\!+3.595559671704640e-5*i} &
\tabnum{\!+6.872261543040661e-4*i} &
\tabnum{-1.447010196851563e-1*i} &
\tabnum{\!+2.609108724913395e-1*i} \\ \midrule
\tworow{$\epsilon^{2}$} &
\tabnum{-1.721862393547420e-4} & 
\tabnum{-7.247299398344942e-4} & 
\tabnum{\!+1.881565035678200e0 } & 
\tabnum{\!+1.544162064068738e0} \\
&
\tabnum{-1.231788432398794e-5*i} &
\tabnum{\!+6.092012063072394e-5*i} &
\tabnum{-4.606206236766448e-3*i} &
\tabnum{\!+1.125263466661532e0*i} \\ \bottomrule \bottomrule 
\end{tabular}
}
\end{center}
\caption{Coefficients of the Laurent expansion in $\epsilon$ for the two-loop non-planar boxes $F^{\boldsymbol{\nu}_{55}}$ in Eq.\,\eqref{eq:2L-box-np-1m-basis} (points $Q_1$ and $Q_2$) and $F^{\boldsymbol{\nu}_{54}}$ in Eq.\,\eqref{eq:2L-box-np-1m-2-basis} (points $P_1$ and $P_2$) at different phase-space points.
  The meaning of each entry is the same as for Table~\ref{tab:1L-triangle-full}.
  More points ($P_3$, $P_4$, $P_5$ and $P_6$) for the second box are shown in Table~\ref{tab:2L-box-np-1m-2}.
}
\label{tab:2L-box-np-1m}
\end{table}

\begin{table}
\begin{center}
\resizebox{15cm}{!}{
\setlength{\tabcolsep}{5pt}
\begin{tabular}{ccccc}
\toprule \toprule 
\tworow{\textbf{target}} &
\tworow{\textbf{$P_3$}} &
\tworow{\textbf{$P_4$}} &
\tworow{\textbf{$P_5$}} &
\tworow{\textbf{$P_6$}} \\ \\ \midrule
\tworow{\textbf{from}} &
\tworow{AMF$^0$} &
\tworow{AMF$^0$, $P_3$} &
\tworow{AMF$^0$} &
\tworow{AMF$^0$, $P_5$} \\ \\ \midrule[\heavyrulewidth]
\tworow{$\epsilon^{0}$} &
\tabnum{\!+2.751593454707949e-1} &
\tabnum{\!+2.506591535092400e-1} &
\tabnum{-2.405260844173886e-1} &
\tworow{\tabnum{-4.831181490833649e-1}} \\
&
\tabnum{-3.815281539209958e-1*i} &
\tabnum{-4.235680397875819e-1*i} &
\tabnum{-5.984661196233730e-3*i} & \\ \midrule
\tworow{$\epsilon^{1}$} &
\tabnum{\!+1.257054227433279e0} & 
\tabnum{\!+1.187415013371159e0} & 
\tabnum{-5.588054474320729e-1} & 
\tworow{\tabnum{-1.396083737425863e0}} \\
&
\tabnum{\!+4.342974425182124e-1*i} &
\tabnum{-5.997939132016630e-1*i} &
\tabnum{-3.250774673987693e-2*i} & \\ \midrule
\tworow{$\epsilon^{2}$} &
\tabnum{\!+2.478546160626464e0} & 
\tabnum{\!+2.372121269639779e0} & 
\tabnum{-1.124284189077083e0} & 
\tworow{\tabnum{-3.146872480560270e0}} \\
&
\tabnum{-2.47049854421279e-1*i} &
\tabnum{-5.961585949177441e-1*i} &
\tabnum{-8.467242364778369e-2*i} &  \\ \bottomrule \bottomrule 
\end{tabular}
}
\end{center}
\caption{
  Coefficients of the Laurent expansion in $\epsilon$ for the two-loop non-planar box $F^{\boldsymbol{\nu}_{54}}$ in Eq.\,\eqref{eq:2L-box-np-1m-2-basis} at different phase-space points.
  The meaning of each entry is the same as for Table~\ref{tab:1L-triangle-full}.
}
\label{tab:2L-box-np-1m-2}
\end{table}

\paragraph{2nd box.}
The box in the right panel of Figure~\ref{fig:2L-box-np-1m-cuts} has inverse propagators

the inverse propagators are
\begin{equation}
  \allowdisplaybreaks
  \begin{aligned}
    D_1 &= k_2^2, &\quad
    D_2 &= (k_2 - p_2)^2 , \\
    D_3 &= (k_2 - k_1 - p_2 - p_3)^2 , &\quad
    D_4 &= k_1^2 - m^2 , \\
    D_5 &= (k_1 - p_1)^2 - m^2 , &\quad
    D_6 &= (k_2 - k_1 - p_2)^2 - m^2 , \\
    D_7 &= (k_2 - k_1 + p_1)^2 - m^2 , &\quad
    D_8 &= (k_1 + p_3)^2 , \\
    D_9 &= (k_1 + p_2)^2.
  \end{aligned}
\end{equation}
The basis $I_i = F^{\boldsymbol{\nu}_i}$ consists of 54 MIs:

{
\small
\allowdisplaybreaks
  \begin{align*}
    \label{eq:2L-box-np-1m-2-basis}
    \boldsymbol{\nu}_1    &= (0, 0, 1, 1, 0, 0, 0, 0, 0) , &\!\!
    \boldsymbol{\nu}_2    &= (1, 1, 1, 0, 0, 0, 0, 0, 0) , &\!\!
    \boldsymbol{\nu}_3    &= (1, -1, 0, 1, 1, 0, 0, 0, 0) , \\
    \boldsymbol{\nu}_4    &= (1, 0, 0, 1, 1, 0, 0, 0, 0) , &\!\!
    \boldsymbol{\nu}_5    &= (-2, 0, 1, 1, 0, 0, 1, 0, 0) , &\!\!
    \boldsymbol{\nu}_6    &= (0, 0, 1, 1, 0, 0, 1, 0, 0) , \\
    \boldsymbol{\nu}_7    &= (0, 0, 0, 1, 1, 0, 1, 0, 0) , &\!\!
    \boldsymbol{\nu}_8    &= (-2, 0, 0, 0, 1, 1, 1, 0, 0) , &\!\!
    \boldsymbol{\nu}_9    &= (0, 0, 0, 0, 1, 1, 1, 0, 0) , \\
    \boldsymbol{\nu}_{10} &= (1, 1, 1, 1, -1, 0, 0, 0, 0) , &\!\!
    \boldsymbol{\nu}_{11} &= (1, 1, 1, 1, 0, 0, 0, 0, 0) , &\!\!
    \boldsymbol{\nu}_{12} &= (1, 0, 1, 1, 1, 0, 0, 0, 0) , \\
    \boldsymbol{\nu}_{13} &= (1, 1, 1, 0, 0, 0, 1, 0, 0) , &\!\!
    \boldsymbol{\nu}_{14} &= (1, 0, 1, 1, 0, 0, 1, 0, 0) , &\!\!
    \boldsymbol{\nu}_{15} &= (1, -1, 0, 1, 1, 0, 1, 0, 0) , \\
    \boldsymbol{\nu}_{16} &= (1, 0, 0, 1, 1, 0, 1, 0, 0) , &\!\!
    \boldsymbol{\nu}_{17} &= (0, 0, 1, 1, 1, 0, 1, 0, 0) , &\!\!
    \boldsymbol{\nu}_{18} &= (1, 0, 0, 0, 1, 1, 1, 0, 0) , \\
    \boldsymbol{\nu}_{19} &= (0, 0, 1, 0, 1, 1, 1, 0, 0) , &\!\!
    \boldsymbol{\nu}_{20} &= (1, 1, 1, 1, 1, 0, 0, 0, 0) , &\!\!
    \boldsymbol{\nu}_{21} &= (1, 0, 1, 1, 1, 1, 0, 0, 0) , \\
    \boldsymbol{\nu}_{22} &= (1, 1, 1, 1, -2, 0, 1, 0, 0) , &\!\!
    \boldsymbol{\nu}_{23} &= (1, 1, 1, 1, -1, 0, 1, 0, 0) , &\!\!
    \boldsymbol{\nu}_{24} &= (1, 1, 1, 1, 0, 0, 1, 0, 0) , \\
    \boldsymbol{\nu}_{25} &= (1, -1, 1, 1, 1, 0, 1, 0, 0) , &\!\!
    \boldsymbol{\nu}_{26} &= (1, 0, 1, 1, 1, -1, 1, 0, 0) , &\!\!
    \boldsymbol{\nu}_{27} &= (1, 0, 1, 1, 1, 0, 1, -1, 0) , \\
    \boldsymbol{\nu}_{28} &= (1, 0, 1, 1, 1, 0, 1, 0, 0) , &\!\!
    \boldsymbol{\nu}_{29} &= (1, 1, -2, 0, 1, 1, 1, 0, 0) , &\!\!
    \boldsymbol{\nu}_{30} &= (1, 1, -1, 0, 1, 1, 1, 0, 0) , \\
    \boldsymbol{\nu}_{31} &= (1, 1, 0, 0, 1, 1, 1, 0, 0) , &\!\!
    \boldsymbol{\nu}_{32} &= (-1, 1, 1, 0, 1, 1, 1, 0, 0) , &\!\!
    \boldsymbol{\nu}_{33} &= (0, 1, 1, -1, 1, 1, 1, 0, 0) , \\
    \boldsymbol{\nu}_{34} &= (0, 1, 1, 0, 1, 1, 1, -1, 0) , &\!\!
    \boldsymbol{\nu}_{35} &= (0, 1, 1, 0, 1, 1, 1, 0, 0) , &\!\!
    \boldsymbol{\nu}_{36} &= (-2, 0, 1, 1, 1, 1, 1, 0, 0) , \\
    \boldsymbol{\nu}_{37} &= (-1, -1, 1, 1, 1, 1, 1, 0, 0) , &\!\!
    \boldsymbol{\nu}_{38} &= (-1, 0, 1, 1, 1, 1, 1, 0, 0) , &\!\!
    \boldsymbol{\nu}_{39} &= (0, 0, 1, 1, 1, 1, 1, 0, 0) , \\
    \boldsymbol{\nu}_{40} &= (1, 1, 1, 1, 1, 1, -2, 0, 0) , &\!\!
    \boldsymbol{\nu}_{41} &= (1, 1, 1, 1, 1, 1, 0, 0, 0) , &\!\!
    \boldsymbol{\nu}_{42} &= (1, 1, 1, 1, 1, -1, 1, 0, 0) , \\
    \boldsymbol{\nu}_{43} &= (1, 1, 1, 1, 1, 0, 1, 0, 0) , &\!\!
    \boldsymbol{\nu}_{44} &= (1, 1, 1, -1, 1, 1, 1, 0, 0) , &\!\!
    \boldsymbol{\nu}_{45} &= (1, 1, 1, 0, 1, 1, 1, 0, 0) , \\
    \boldsymbol{\nu}_{46} &= (1, -2, 1, 1, 1, 1, 1, 0, 0) , &\!\!
    \boldsymbol{\nu}_{47} &= (1, -1, 1, 1, 1, 1, 1, 0, 0) , &\!\!
    \boldsymbol{\nu}_{48} &= (1, 0, 1, 1, 1, 1, 1, -1, 0) , \\
    \boldsymbol{\nu}_{49} &= (1, 0, 1, 1, 1, 1, 1, 0, 0) , &\!\!
    \boldsymbol{\nu}_{50} &= (1, 1, 1, 1, 1, 1, 1, -3, 0) , &\!\!
    \boldsymbol{\nu}_{51} &= (1, 1, 1, 1, 1, 1, 1, -2, 0) , \\
    \boldsymbol{\nu}_{52} &= (1, 1, 1, 1, 1, 1, 1, -1, -1) , &\!\!
    \boldsymbol{\nu}_{53} &= (1, 1, 1, 1, 1, 1, 1, -1, 0) , &\!\!
    \boldsymbol{\nu}_{54} &= (1, 1, 1, 1, 1, 1, 1, 0, 0) \numthis .
  \end{align*}
}

We consider three propagations that cross separately the three branch points associated with $z_1 = s - 4 m^2$, $z_2 = t - 9 m^2$, $z_3 = -s - t + 7 m^{2}$.
In particular, we cross:
\begin{itemize}
\item $z_1$ going from point $P_1$ with $s = 3$, $t = 2$, $m^2 = 1$ ($z_1 < 0 $) to $P_2$ with $s = 5$ and same values for $t$ and $m^2$ ($z_1 > 0 $);
\end{itemize}
\begin{itemize}
\item $z_2$ propagating from  $P_3$ with $s = 2$, $t = 8$, $m^2 = 1$ ($z_2 < 0 $) to $P_4$ with $t = 10$ and same values for $s$ and $m^2$ ($z_2 > 0 $);
\end{itemize}
\begin{itemize}
\item $z_3$ moving from point $P_5$ with $s = -3$, $t = -5$, $m^2 = 1$ ($z_3 > 0 $) to $P_6$ with $s = -1$, $t = -3$ and same value for $m^2$ ($z_3 < 0 $).
\end{itemize}

\noindent
Internal consistency is obtained in all points using AMF$^0$ with an extended basis of 89 MIs.
The numerical results for the two-loop non-planar box $F^{\boldsymbol{\nu}_{54}}$ are shown in Tables~\ref{tab:2L-box-np-1m} and~\ref{tab:2L-box-np-1m-2}.

\section{Conclusion}
\label{sec:conclusion}
In this paper, we have presented \Line{}, a novel open-source code available at
\begin{center}
  \url{https://github.com/line-git/line.git}
\end{center}
for the numerical evaluation of Feynman integrals through the solution of differential equations via series expansions. We have detailed the methods implemented within the framework and showcased its capabilities through a suite of illustrative examples.

\Line{} is designed as a publicly available, modular, and efficient tool, making it suitable for widespread use in phenomenological applications. The code can function as a standalone tool to propagate boundary values for problems involving (in principle) any number of loops. When boundary values are not readily available, the auxiliary-mass flow method and its automated boundary determination technique have been implemented for problems up to two loops.

The challenges posed by boundary conditions remain an important area of research. Our code lays the groundwork for addressing these challenges by offering multiple development possibilities. For instance, a recursive implementation of the AMFlow method could extend \Line{} applicability to higher-loop problems. Additionally, a systematic study of the EBR method to automate boundary determination at all $\epsilon$ orders, exploiting the singular structure of DEs, represents a promising avenue for future work.

To promote accessibility and usability, we have made an effort to rely exclusively on well-maintained, open-source libraries. This ensures compatibility and efficiency when deploying \Line{} on high-performance computing clusters.

This release represents the first version of \Line{}, where our primary focus has been on verifying the correctness of the results. However, we recognize numerous opportunities to improve computational efficiency. Thanks to the modular structure of the code, such enhancements can be implemented with relative ease in future iterations.

In conclusion, \Line{} provides a robust foundation for solving complex problems in perturbative quantum field theory.
We hope that its open-source nature and modular architecture can make it a valuable tool for the community.


\appendix

\section{Matrix normalization around regular singular points}
\label{sec:matr-norm}
The transformation of the DE matrix $\mathbb{A}$ to \textit{Fuchsian normal form} can be carried out exploiting its block-diagonal structure.
The blocks on the diagonal of $\mathbb{A}$ can be used to identity a block grid across the whole matrix.
To see this, let $(i_r,i_r)$ be the row and column indices of the top-left matrix element of the $r$-th block and $L_r$ its dimension, with $r = 1, \dots, B$, where $B$ is the number of blocks on the diagonal.
We can label as $\mathbb{A}^{(r,s)}$ the $L_r \times L_s$ block whose top-left element indices are $(i_r, i_s)$.
In this way, $\mathbb{A}^{(r,s)}$ represents the  block extracted from the $r$-th row and $s$-th column of the block grid composing $\mathbb{A}$, while $\mathbb{A}^{(r,r)}$ denotes the $r$-th block on the diagonal.

The DE matrix $\mathbb{A}$ can be transformed to Fuchsian normal form with the following two steps:
\begin{enumerate}
\item
  The blocks on the diagonal $\mathbb{A}^{(r,r)}$ are transformed to Fuchsian normal form one by one, each block being first put in Fuchsian form and then normalized. We refer to this step as diagonal normalization.

\item
  The off-diagonal blocks are transformed to Fuchsian form too.
\end{enumerate}

\paragraph{1. Diagonal normalization.}
The diagonal block $\mathbb{A}^{(r,r)}$ can be transformed to Fuchsian form following the algorithm in \cite{Lee:2014ioa}.
The output are the transformed block $\widetilde{\mathbb{A}}^{(r, r)}_f$ with Laurent expansion
\begin{equation}
    \widetilde{\mathbb{A}}^{(r,r)}(\eta) = \frac{1}{\eta} \sum_{k=0}^\infty\widetilde{\mathbb{A}}^{(r,r)}_k \eta^k \,,
\end{equation}
and the corresponding transformation matrix $\mathbb{T}^{(r,r)}(\eta)$, whose elements are rational functions in the line parameter $\eta$.

The leading order $\widetilde{\mathbb{A}}^{(r, r)}_0$ is then put in Jordan form and the corresponding ($\eta$-independent) transformation $\mathbb{T}^{(r, r)}_J$ is cumulated with the previous one according to
\begin{align*}
  \mathbb{T}^{(r,r)}(\eta) &\rightarrow \mathbb{T}^{(r,r)}(\eta) \mathbb{T}^{(r,r)}_J \,, \\
  \widetilde{\mathbb{A}}^{(r,r)}(\eta) &\rightarrow \mathbb{T}^{(r,r) -1}_J \widetilde{\mathbb{A}}^{(r,r)}(\eta) \mathbb{T}^{(r,r)}_J \,.
\end{align*}
If all the eigenvalues of leading order $\widetilde{\mathbb{A}}^{(r,r)}_0$ have real part in $[0, 1[$, the block is already normalized and no further action is required.
Otherwise, \textit{shearing transformations} can be implemented to shift the eigenvalue of any Jordan chain by $\pm 1$ so that its real part progressively moves towards $[0, 1[$.

The shearing transformation matrix is
\begin{equation}
  \label{def:shear-transf}
  \mathbb{T}^{(r,r)}_{\text{sh}}(\eta) \equiv \text{diag}(1, \dots, 1, \eta^{-s}, \dots, \eta^{-s}, 1, \dots, 1)\,, \quad \text{shift by}\, s = \pm 1 \,,
\end{equation}
where $\eta^{-s}$ is placed at the indices corresponding to the Jordan chain to be shifted in $\widetilde{\mathbb{A}}^{(r,r)}_0$.
When transforming the block using
\begin{equation}
  \widetilde{\mathbb{A}}^{(r,r)}(\eta) \rightarrow \mathbb{T}^{(r,r) -1}_{\text{sh}}(\eta) \widetilde{\mathbb{A}}^{(r,r)}(\eta) \mathbb{T}^{(r,r)}_{\text{sh}}(\eta) - \mathbb{T}^{(r,r)-1}_{\text{sh}}(\eta) \frac{d}{d\eta}\mathbb{T}^{(r,r)}_{\text{sh}}(\eta) \,,
\end{equation}
 the actual shift of the eigenvalue comes from the derivative contribution
\begin{equation}
  - \mathbb{T}^{(r,r)-1}_{\text{sh}} \frac{d}{d\eta}\mathbb{T}^{(r,r)}_{\text{sh}} = \text{diag} (0, \dots, 0, s\,\eta^{-1}, \dots, s\,\eta^{-1}, 0, \dots, 0) \,,
\end{equation}
which acts on the diagonal of the leading order $\widetilde{\mathbb{A}}^{(r,r)}_0$.
On the other hand, the contribution $\mathbb{T}^{(r,r) -1}_{\text{sh}}(\eta) \widetilde{\mathbb{A}}^{(r,r)}(\eta) \mathbb{T}^{(r,r)}_{\text{sh}}(\eta)$ simply multiplies the columns corresponding to the Jordan chain by $\eta^{-s}$ and its rows by $\eta^s$.
Therefore, in $\widetilde{\mathbb{A}}^{(r,r)}_0$ the block corresponding to the chain remains unchanged, while some matrix elements from the next-to-leading order $\widetilde{\mathbb{A}}^{(r,r)}_1$ appears off-block in the leading order due to the $\eta^{-1}$ factor.
The leading order has to be transformed again to Jordan form
and the overall result is that the Jordan chain is shifted as a whole, preserving its structure.
The transformation matrix $\mathbb{T}^{(r,r)}$ is updated according to
\begin{equation}
  \mathbb{T}^{(r,r)}(\eta) \rightarrow \mathbb{T}^{(r,r)}(\eta) \mathbb{T}^{(r,r)}_{\text{sh}}(\eta) \mathbb{T}^{(r,r)}_J \,.
\end{equation}

Additional shearing and Jordan transformations can be sequentially applied until all eigenvalues have their real part in $[0, 1[$.
Note that, if necessary, multiple Jordan chains (even with different eigenvalues) can be treated together with one shearing transformation simply placing, in Eq.\,\eqref{def:shear-transf}, a shifting term at the indices corresponding to every chain.

The final output of this procedure are the transformed block $\widetilde{\mathbb{A}}^{(r,r)}(\eta)$ in Fuchsian normal form and the cumulated transformation matrix $\mathbb{T}^{(r,r)}(\eta)$ such that
\begin{equation}
  \widetilde{\mathbb{A}}^{(r,r)}(\eta) = \mathbb{T}^{(r,r) -1}(\eta) \mathbb{A}^{(r,r)}(\eta) \mathbb{T}^{(r,r)}(\eta) - \mathbb{T}^{(r,r)-1}(\eta) \frac{d}{d\eta}\mathbb{T}^{(r,r)}(\eta) \,.
\end{equation}




By normalizing the blocks on the diagonal, we built a block-diagonal transformation matrix
\begin{equation}
  \mathbb{T}(\eta) \equiv \text{diag} (\mathbb{T}^{(1,1)}, \mathbb{T}^{(2,2)}, \dots, \mathbb{T}^{(B,B)}) \,.
\end{equation}
Of course, when acting on $\mathbb{A}(\eta)$, such matrix also changes the off-diagonal blocks to
\begin{equation}
  \widetilde{\mathbb{A}}^{(r,s)}(\eta) = \mathbb{T}^{(r,r) -1}(\eta) \mathbb{A}^{(r,s)}(\eta) \mathbb{T}^{(s,s)}(\eta) \,,\quad s < r \,.
\end{equation}

\paragraph{2. Off-diagonal fuchsianization.}
So far we put the blocks on the diagonal in Fuchsian normal form, but the off-diagonal blocks are still, in general, non-Fuchsian.
Therefore, we look for a transformation that changes these blocks only, while preserving the ones on the diagonal.
Such a property is verified by any block-lower triangular  matrix $\mathbb{T}_{\off}$ with identity matrices on the diagonal:
\begin{align}
  \mathbb{T}_{\off}
  &=\left(\begin{array}{ccN{1cm}N{1.5cm}N{1.5cm}N{1.5cm}}
    \mathbbm{1} & 0 & 0 & 0 & \dots & 0 \\
    \mathbb{T}^{(2,1)}_{\off} & \mathbbm{1} & 0 & 0 & \dots & 0 \\
    \mathbb{T}^{(3,1)}_{\off} & \mathbb{T}^{(3,2)}_{\off} & \mathbbm{1} & 0 & \dots & 0 \\[0.3cm]
    \vdots & \vdots & \ddots & \ddots & \ddots & \vdots \\[0.3cm]
    \mathbb{T}^{(B-1,1)}_{\off} & \mathbb{T}^{(B-1,2)}_{\off} & \dots & \mathbb{T}^{(B-1,B-1)}_{\off} & \mathbbm{1} & 0 \\
    \mathbb{T}^{(B,1)}_{\off} & \mathbb{T}^{(B,2)}_{\off} & \dots & \mathbb{T}^{(B,B-2)}_{\off}  & \mathbb{T}^{(B,B-1)}_{\off} & \mathbbm{1}
  \end{array}\right) \,.
\end{align}
This matrix can be decomposed as
\begin{align*}
  \label{eq:off-diag-fuch-order}
  \mathbb{T}_{\off} &=
  \begin{pmatrix}
    \mathbbm{1} & 0 & 0 & 0 & \dots & 0 \\
    \mathbb{T}^{(2,1)}_{\off} & \mathbbm{1} & 0 & 0 & \dots & 0 \\
    0 & 0 & \mathbbm{1} & 0 & \dots & 0 \\
    \vdots & \vdots & \ddots & \ddots & \ddots & \vdots \\
    0 & 0 & \dots & 0 & \mathbbm{1} & 0 \\
    0 & 0 & \dots & 0 & 0 & \mathbbm{1} \\
  \end{pmatrix}\cdot
  \begin{pmatrix}
    \mathbbm{1} & 0 & 0 & 0 & \dots & 0 \\
    0 & \mathbbm{1} & 0 & 0 & \dots & 0 \\
    0 & \mathbb{T}^{(3,2)}_{\off} & \mathbbm{1} & 0 & \dots & 0 \\
    \vdots & \vdots & \ddots & \ddots & \ddots & \vdots \\
    0 & 0 & \dots & 0 & \mathbbm{1} & 0 \\
    0 & 0 & \dots & 0 & 0 & \mathbbm{1} \\
  \end{pmatrix}\cdot
  \begin{pmatrix}
    \mathbbm{1} & 0 & 0 & 0 & \dots & 0 \\
    0 & \mathbbm{1} & 0 & 0 & \dots & 0 \\
    \mathbb{T}^{(3,1)}_{\off} & 0 & \mathbbm{1} & 0 & \dots & 0 \\
    \vdots & \vdots & \ddots & \ddots & \ddots & \vdots \\
    0 & 0 & \dots & 0 & \mathbbm{1} & 0 \\
    0 & 0 & \dots & 0 & 0 & \mathbbm{1} \\
  \end{pmatrix}\cdot
  \dots \\[0.3cm]
  &\equiv \mathbb{T}_{\off}[2,1] \,\mathbb{T}_{\off}[3,2] \,\mathbb{T}_{\off}[3,1] \dots \,, \numthis
\end{align*}
where we indicate with $\mathbb{T}_{\off}[r,s]$ the matrix with identity matrices on the diagonal and whose only non-zero off-diagonal block is $\mathbb{T}^{(r,s)}_\off$.
In the above factorization, the order of the factors can be chosen arbitrarily, however we select the one
where the only non-zero off-diagonal block goes through each row $r$ starting from the block $(r,r-1)$ and moving towards the block $(r,1)$, then proceeding with the next row and so on.
We do so because the single transformation $\mathbb{T}_{\off}[r,s]$
acts on $\widetilde{\mathbb{A}}$ through
\begin{equation}
  \widetilde{\mathbb{A}} \rightarrow \mathbb{T}^{-1}_{\off}[r,s] \widetilde{\mathbb{A}} \mathbb{T}_{\off}[r,s] - \mathbb{T}^{-1}_{\off}[r,s] \frac{d}{d\eta} \mathbb{T}_{\off}[r,s]
\end{equation}
by changing the $(r,s)$ block, the $r$-th row of blocks from $(r,1)$ up to $(r, s-1)$ and the $s$-th column of blocks from $(r+1,s)$ down to $(B,s)$.
In particular, we have
\begin{equation}
  \label{eq:off-diag-transf}
  \widetilde{\mathbb{A}}^{(h,k)} \rightarrow
  \left\{
  \begin{array}{lc}
   \widetilde{\mathbb{A}}^{(r,s)} - \mathbb{T}^{(r,s)}_{\off} \widetilde{\mathbb{A}}^{(s,s)} + \widetilde{\mathbb{A}}^{(r,r)} \mathbb{T}^{(r,s)}_{\off} - \frac{d}{d\eta} \mathbb{T}^{(r,s)}_{\off}  &\quad h = r,\,\, k = s \\
   \widetilde{\mathbb{A}}^{(r,k)} - \mathbb{T}^{(r,s)} \widetilde{\mathbb{A}}^{(s,k)}  &\quad h = r,\,\, k < s \\
   \widetilde{\mathbb{A}}^{(h,s)} + \widetilde{\mathbb{A}}^{(h,r)}  \mathbb{T}^{(r,s)} &\quad h > r,\,\, k = s \\
   \widetilde{\mathbb{A}}^{(h,k)} &\quad \text{otherwise} \,.
  \end{array}
  \right.
\end{equation}
A suitable block $\mathbb{T}^{(r,s)}_{\off}$ can be found in order to lower the Poincaré rank of $\widetilde{\mathbb{A}}^{(r,s)}$ down to zero, as shown below.
However, while doing so we see from Eq.\,\eqref{eq:off-diag-transf} that other blocks on the sub-row and the sub-column of $(r,s)$ are changed too.
Therefore, in order not to spoil the fuchsianization of $\widetilde{\mathbb{A}}^{(r,s)}$ in later iterations, we start by reducing the rank of $\widetilde{\mathbb{A}}^{(1,1)}$ and then we proceed with $\widetilde{\mathbb{A}}^{(2,1)}$, $\widetilde{\mathbb{A}}^{(3,2)}$,  $\widetilde{\mathbb{A}}^{(3,1)}$, \dots, thus explaining the choice made for the order of the factors in Eq.\,\eqref{eq:off-diag-fuch-order}.

To lower the Poincaré rank $p$ of $\widetilde{\mathbb{A}}^{(r,s)}$,
we look for a block $\mathbb{T}^{(r,s)}_{\off}$ in the form
\begin{equation}
  \mathbb{T}^{(r,s)}_{\off} = \frac{\mathbb{G}^{(r,s)}}{\eta^p}, \quad \mathbb{G}^{(r,s)} = \text{const} \,.
\end{equation}
By substituting into Eq.\,\eqref{eq:off-diag-transf} and imposing the cancellation of the leading order $\eta^{p+1}$ we obtain
\begin{equation}
  \label{eq:off-diag-fuch-lin-sys}
  0 = \widetilde{\mathbb{A}}^{(r,s)}_0 - \mathbb{G}^{(r,s)} \widetilde{\mathbb{A}}^{(s,s)}_0 + \widetilde{\mathbb{A}}^{(r,r)}_0 \mathbb{G}^{(r,s)} + p\, \mathbb{G}^{(r,s)} \,,
\end{equation}
which, element by element, gives us a system of linear equations for the $L_r \times L_s$ unknown matrix elements of $\mathbb{G}^{(r,s)}$.
By construction, the matrix $\mathbb{T}_{\off}^{(r,s)}$ resulting from the solution of the system lowers the Poincaré rank $p$ of $\widetilde{\mathbb{A}}^{(r,s)}$.

We can iterate this procedure until $p = 0$, and then proceed with the next off-diagonal block of $\mathbb{A}$ in the order specified above.



\section{Implementation details}
\label{sec:implementation}
\subsection{Representation of rational functions}
\label{sec:rat-func}
\Line{} relies on the manipulation of rational functions for many tasks, such as building the DEs along the phase-space line, finding poles, transforming the system to a Fuchsian form, or solving recurrence relations.
These steps require to implement a representation of rational functions that allows to efficiently perform operations such as sums, products, shifts or expansions around poles.

Consider a rational function with numerator~$N(x)$ of degree~$d_N$ and denominator~$D(x)$ of degree~$d_D$,
\begin{equation}
  \label{eq:1}
  f(\eta) = \frac{N(\eta)}{D(\eta)} = \frac{\sum_{j=0}^{d_N} a_j \eta^j}{\sum_{k=0}^{d_D} b_k \eta^k} \quad.
\end{equation}

In \Line{} the numerator is stored through the~$d_N + 1$ coefficients~$(a_j)_{j = 0, \dots, d_N}$, while the denominator is represented by its roots~$(r_k)_k$, thus referring to its factorized form
\begin{equation}
  \label{eq:3}
  D(\eta) = \prod_k(\eta - r_k)^{m_k} \quad,
\end{equation}
where $m_k$ is the multiplicity of the root $r_k$. Here the choice~$b_{d_D} = 1$ is implied, i.e. the denominator~$D(x)$ is always made \textit{monic} by absorbing~$b_{d_D}$ in a redefinition of the other coefficients~$a_j \rightarrow a_j/b_{d_D}$,~$b_k \rightarrow b_k/b_{d_D}$.
Both coefficients and roots are stored at arbitrary precision with~\texttt{mpc\_t}, the latter being computed numerically starting from the coefficients of the denominator.

The above representation by roots becomes very useful when considering that roots in the denominators of Feynman Integrals DEs typically appear multiple times across different matrix elements.
In \Line{} we exploit this aspect by storing each root only once, updating a global list of unique roots every time a new one is found.
Each root of this list is assigned an unique integer label so that, for every denominator, only the label of the roots (and the associated multiplicities) are stored. The advantage is twofold, reducing the memory footprint (integers labels use less memory than arbitrary precision roots) and allowing for faster execution of certain operations. For instance, the computation of the LCM of two denominators reduces to simply comparing integer labels to determine which roots appear and with what multiplicity. Also, to shift all the denominators around a point~$\eta_s$ (which is useful to recenter the DEs so that we always solve around the origin) all it takes is shifting only once every root in the global list,~$r_k \rightarrow r_k-\eta_s$, while the labels associated with every denominator remain unchanged.

\subsection{Mathematical expressions}\label{sec:math-expr}
\Line{} accepts in input files containing symbolic mathematical expressions.
The elements of the DE matrices, for instance, are rational functions depending on kinematic invariants and the space-time dimension.
These expression have to be processed to go from the input string to the actual representation shown in Section~\ref{sec:rat-func}.

The first step to do so consists in parsing the string of the mathematical expression, encoding it into an internal representation of the associated \textit{expression tree}.
Within such a tree, nodes may represent operations, symbols, or numbers.
Operation nodes have child nodes as their operands, while symbols and numbers serve as terminal nodes with no children.

To simplify memory management for node insertion and deletion, child nodes are arranged in a \textit{linked list} where each node points to one sibling, while only the connection between the parent and its first child is maintained.
In case of commutative operations, the choice of the first child is of course arbitrary.

The implementation within \Line{} of this mathematical structure includes basic operations that are useful to express the DE matrix elements as rational functions in the phase-space line parameter, such as the expansion of product or the extraction of polynomial coefficients.

The actual conversion of an expression tree into its representation in terms of numerator coefficients and denominator roots is performed by a decoding routine that recursively navigates the tree, updating the global list of roots as soon as new denominators are analyzed.

\pagebreak

\bibliographystyle{JHEP}
\bibliography{biblio}

\end{document}